\documentclass[11pt,a4paper]{article}

\usepackage[T1]{fontenc}
\usepackage[utf8]{inputenc}
\usepackage{lmodern}
\ifdefined\pdfglyphtounicode
  \IfFileExists{glyphtounicode-cmr.tex}{\input{glyphtounicode-cmr.tex}}{}
  \pdfglyphtounicode{summationdisplay}{2211}\pdfglyphtounicode{summationtext}{2211}
  \pdfglyphtounicode{productdisplay}{220F}\pdfglyphtounicode{producttext}{220F}
  \pdfglyphtounicode{intersectiondisplay}{22C2}\pdfglyphtounicode{intersectiontext}{22C2}
  \pdfglyphtounicode{tildewide}{0303}\pdfglyphtounicode{tildewider}{0303}\pdfglyphtounicode{tildewidest}{0303}
\fi
\usepackage[a4paper,margin=25mm,headheight=15pt]{geometry}
\usepackage{amsmath,amssymb,amsthm,mathtools,bm}
\usepackage{microtype,booktabs,array,multirow}
\usepackage{graphicx,xcolor,tikz,subcaption}
\usepackage{flafter}
\usepackage[section]{placeins}
\usetikzlibrary{calc,arrows.meta,positioning,patterns}
\usepackage{enumitem}
\usepackage{hyperref}
\usepackage[capitalise,noabbrev]{cleveref}
\usepackage{fancyhdr}
\usepackage[noblocks]{authblk}

\definecolor{navy}{HTML}{18364C}
\definecolor{teal}{HTML}{007F91}
\definecolor{orange}{HTML}{CB7128}
\definecolor{purple}{HTML}{76549B}

\hypersetup{
  colorlinks=true,
  linkcolor=navy,
  citecolor=teal,
  urlcolor=teal
}

\ifdefined\pdfomitinfodict
\fi
\ifdefined\pdfsuppressptexinfo
\fi
\ifdefined\pdftrailerid
  \pdftrailerid{}
\fi

\setlist[enumerate]{itemsep=3pt,topsep=5pt}
\allowdisplaybreaks[1]

\newtheorem{theorem}{Theorem}
\newtheorem{proposition}[theorem]{Proposition}
\newtheorem{lemma}[theorem]{Lemma}

\theoremstyle{definition}

\theoremstyle{remark}

\newcommand{\ind}[1]{\mathbf{1}[#1]}
\newcommand{\R}{\mathbb{R}}
\newcommand{\Lloc}{\mathcal{L}}
\newcommand{\Naff}{\mathcal{N}}
\newcommand{\U}{\mathcal{U}}
\newcommand{\Ccal}{\mathcal{C}}
\newcommand{\aff}{\operatorname{aff}}
\newcommand{\conv}{\operatorname{conv}}

\newcommand{\sat}{\mathrm{sat}}
\newcommand{\pa}{p_A}
\newcommand{\pb}{p_B}

\tikzset{
  every picture/.style={line cap=round,line join=round},
  dot/.style={circle,fill=navy,inner sep=2pt},
  >=Latex
}

\title{%
\textbf{How many settings can a tight Bell inequality require?}
}

\author[1,2]{Emmanuel Zambrini Cruzeiro}
\author[3,4]{Nicolas Gisin}

\affil[1]{%
Quantum Physics of Information (QPI) Group,
Instituto de Telecomunicações, Lisbon, Portugal
}

\affil[2]{%
Quantum Information and Quantum Optics (QIQO) Laboratory,
Instituto Superior Técnico, Universidade de Lisboa, Lisbon, Portugal
}

\affil[3]{%
Department of Applied Physics,
University of Geneva, Geneva, Switzerland
}

\affil[4]{%
Constructor University,
28759 Bremen, Germany
}

\date{}

\begin{document}
\maketitle
\thispagestyle{plain}
\begin{abstract}
Consider a bipartite Bell experiment in which Alice has $X\geq2$ measurement settings with $A\geq2$ outcomes each. We prove that a tight Bell inequality can require at most $K=A^X-X(A-1)$ of Bob's settings, and construct facets that require all $K$. The result holds for every finite number $B\geq2$ of Bob outcomes. A local model must reproduce the data for every Bob setting using the same distribution over Alice's deterministic responses. Once Alice's observed marginals are fixed, the compatible distributions over her deterministic responses lie in an affine space of dimension at most $K-1$. Helly's theorem then implies that if every restriction of a nonsignalling behavior to at most $K$ of Bob's settings is local, the full behavior is local. This cutoff is tight: there exist nonlocal nonsignalling behaviors with $K$ Bob settings whose restriction to every smaller subset is local. For two ternary Alice measurements and binary Bob outcomes, complete facet enumerations with five and six Bob settings give ten classes, including positivity. Together with the cutoff $K=5$, these enumerations give a complete classification, up to relabellings and input liftings, for arbitrarily many Bob settings.
\end{abstract}

\section{Introduction}

Bell inequalities distinguish correlations that admit a local hidden-variable model from those that do not~\cite{Bell,CHSH,Brunner}. For a fixed number of measurement settings and outcomes, the local correlations form a polytope. Its facets define the tight inequalities, including positivity constraints. A full description requires every distinct facet. When classifying them, one can instead give one representative of each symmetry class and obtain the remaining facets by relabelling settings and outcomes.

Increasing the number of settings can produce new inequalities. The inequality $I_{3322}$, for example, detects the nonlocality of some two-qubit states for which no CHSH violation is possible~\cite{CollinsGisin}. Complete classifications~\cite{CruzeiroGisin,JesusCruzeiro} and recent sampling results~\cite{Staufenbiel} show how rapidly the number of inequivalent facets can grow even in small Bell scenarios. However, a larger scenario also contains inequalities that were already present in a smaller one. An inequality can simply ignore an added setting. This is an input lifting, and such liftings preserve facets~\cite[Theorem~6]{Pironio}.

Here we fix Alice's measurement scenario and allow Bob to add settings. We ask whether this process eventually stops producing facets that require more of Bob's settings. We prove that it does, and determine the exact stopping point:
\begin{equation}
 K=A^X-X(A-1).
 \label{eq:K}
\end{equation}
Every facet can be written using at most $K$ Bob settings. There are also facets for which all $K$ settings are necessary. Both statements hold for every finite number $B\geq2$ of Bob outcomes.

The case $X=A=2$, for which $K=2$, follows from Pironio's classification: every nontrivial facet in this case is a lifting of CHSH, irrespective of Bob's numbers of settings and outcomes~\cite{PironioCHSH}. Here we determine the exact cutoff for arbitrary $X,A\geq2$.

The upper bound follows by asking whether the correlations for Bob's different settings can be explained by a single local model. In a deterministic local strategy, Alice has a predetermined outcome for each of her $X$ settings. There are $A^X$ possible assignments of these outcomes, and a general local model assigns probabilities to them. These probabilities must reproduce Alice's observed marginals. For each Bob setting, we must also be able to choose his response, conditioned on Alice's predetermined answers, so that the model reproduces the observed Alice--Bob correlations. The same distribution over Alice's answers must work for every Bob setting.

For each Bob setting, the distributions over Alice's answers that allow such a model form a convex set. The full behavior is local precisely when these sets have a common distribution. All these sets lie in a common affine space of dimension at most $A^X-1-X(A-1)=K-1$: the $A^X$ probabilities satisfy normalization and $X(A-1)$ independent constraints fixing Alice's marginals. Helly's theorem then implies that, if no distribution works for all Bob settings, there is already a subset of at most $K$ settings for which none works. Thus every nonlocal nonsignalling behavior remains nonlocal after retaining a suitable subset of at most $K$ Bob settings. We use this locality criterion to show that, whenever Bob has more than $K$ settings, every facet inequality is a lifting of a facet inequality from a scenario with at most $K$ of his settings.

To prove that this bound cannot be improved, we construct tight Bell inequalities for which all $K$ Bob settings are necessary. For each inequality, we establish its local bound and identify the deterministic strategies that saturate it. We prove that their affine span has dimension one less than the local polytope, establishing tightness. We also show that the inequality cannot be rewritten using fewer Bob settings, even after using normalization and nonsignalling identities. When Bob has two outcomes, we give one construction for two Alice settings and another for three or more, each yielding a facet that requires all $K$ of Bob's settings. For each finite $B>2$, we group all nonzero Bob outcomes together and show that the resulting inequality defines a facet of the $B$-outcome local polytope that still requires all $K$ settings.

For two ternary Alice settings and binary Bob outcomes, the cutoff is five. We give the complete list of ten facet classes in the $(2,5,3,2)$ scenario and report a second complete enumeration for $(2,6,3,2)$. Two of the ten classes require all five Bob settings; one is obtained directly from our analytic construction. The theorem explains why adding a sixth setting, or any further settings, cannot produce a facet that is not an input lifting of one of these ten classes.

\section{Bell scenarios and the statement of the result}
\label{sec:scenario}

\subsection{Local correlations and their affine space}

Alice chooses $x\in\{1,\ldots,X\}$ and obtains $a\in\{0,\ldots,A-1\}$. Bob chooses $y\in\{1,\ldots,Y\}$ and obtains $b\in\{0,\ldots,B-1\}$. We denote this scenario by $(X,Y,A,B)$. Its correlations are the conditional probabilities $p(a,b|x,y)$.

Normalization and nonsignalling require
\begin{align}
 \sum_{a,b}p(a,b|x,y)&=1,\\
 \sum_b p(a,b|x,y)&=\pa(a|x),\\
 \sum_a p(a,b|x,y)&=\pb(b|y),
 \label{eq:NS}
\end{align}
where Alice's marginal is independent of $y$ and Bob's marginal is independent of $x$. Let $\Naff^{A,B}_{X,Y}$ denote the affine space defined by these equalities. A point of this affine space is a \emph{nonsignalling behavior} if all its entries $p(a,b|x,y)$ are nonnegative.

A deterministic local strategy is specified by two strings,
\begin{equation}
 \alpha=(\alpha_1,\ldots,\alpha_X)\in\Omega:=\{0,\ldots,A-1\}^X,
 \qquad
 \beta=(\beta_1,\ldots,\beta_Y)\in\{0,\ldots,B-1\}^Y.
\end{equation}
The entries of $\alpha$ are Alice's predetermined answers to all her settings; similarly for $\beta$. The corresponding behavior is
\begin{equation}
 p_{\alpha,\beta}(a,b|x,y)=\ind{a=\alpha_x}\ind{b=\beta_y}.
 \label{eq:det}
\end{equation}
The local polytope is
\begin{equation}
 \Lloc^{A,B}_{X,Y}=\conv\{p_{\alpha,\beta}\}.
 \label{eq:localpoly}
\end{equation}
Thus $p$ is local if and only if there are numbers $\mu_{\alpha\beta}\geq0$, summing to one, such that
\begin{equation}
 p(a,b|x,y)=\sum_{\alpha,\beta}\mu_{\alpha\beta}
 \ind{a=\alpha_x}\ind{b=\beta_y}.
 \label{eq:localmixture}
\end{equation}
This is the standard formulation of locality through a joint distribution over predetermined responses~\cite{Fine}. The strings describe a hidden-variable model; they do not require the different settings to be measured together.

Independent coordinates for $\Naff^{A,B}_{X,Y}$ are
\begin{equation}
 \pa(a|x),\qquad \pb(b|y),\qquad p(a,b|x,y),
 \qquad a=1,\ldots,A-1,\quad b=1,\ldots,B-1.
 \label{eq:coords}
\end{equation}
The probabilities with one or both outcomes zero follow from
\begin{align}
 p(a,0|x,y)&=\pa(a|x)-\sum_{b=1}^{B-1}p(a,b|x,y),\\
 p(0,b|x,y)&=\pb(b|y)-\sum_{a=1}^{A-1}p(a,b|x,y),\\
 p(0,0|x,y)&=1-\sum_{a=1}^{A-1}\pa(a|x)-\sum_{b=1}^{B-1}\pb(b|y)
 +\sum_{a=1}^{A-1}\sum_{b=1}^{B-1}p(a,b|x,y).
\end{align}
Consequently,
\begin{equation}
 D:=\dim\Naff^{A,B}_{X,Y}
 =X(A-1)+Y(B-1)+XY(A-1)(B-1).
 \label{eq:dimension}
\end{equation}

The local polytope has dimension $D$ in the normalized nonsignalling affine space $\Naff^{A,B}_{X,Y}$~\cite[Theorem~1]{Pironio}. A \emph{Bell functional}, also called a \emph{Bell expression}, assigns each behavior $p$ a number $\mathcal I(p)$ by taking a weighted sum of its probabilities and allowing for a constant term. If $L$ is the maximum of $\mathcal I(p)$ over local behaviors, then $\mathcal I(p)\leq L$ is the corresponding \emph{Bell inequality}. We call $L$ its \emph{local bound}. The inequality is \emph{tight} if the deterministic local strategies saturating it, $\mathcal I(p_{\alpha,\beta})=L$, have an affine span of dimension $D-1$. Their convex hull is then a facet of the local polytope.

\subsection{Removing settings and input liftings}

For a nonempty subset $S\subseteq\{1,\ldots,Y\}$, let $\pi_S(p)$ be the restriction of $p$ to Bob settings in $S$. An inequality on the restricted scenario becomes an inequality on the full scenario by writing
\begin{equation}
 f(\pi_S(p))\geq0.
 \label{eq:lifting}
\end{equation}
It simply ignores the settings outside $S$. We call this an input lifting.

Whether a setting is used is a property of the affine functional, not of a particular redundant array of probability coefficients. Normalization and nonsignalling can change that array without changing the functional. We say that a Bob setting is \emph{indispensable} if the functional cannot be written using only the other settings, even after using these identities.

Let $Y_{\sat}(X;A,B)$ be the least integer $K$ such that, for every $Y>K$, each facet of $\Lloc^{A,B}_{X,Y}$ is an input lifting of a facet with at most $K$ Bob settings.

\par\noindent\begin{minipage}{\textwidth}
\begin{theorem}[Exact setting saturation]
\label{thm:main}
For all finite $X,A,B\geq2$,
\begin{equation}
 Y_{\sat}(X;A,B)=A^X-X(A-1).
 \label{eq:main}
\end{equation}
More precisely, writing $K=A^X-X(A-1)$:
\begin{enumerate}
 \item Every facet in any $(X,Y,A,B)$ scenario is an input lifting of a facet involving at most $K$ Bob settings.
 \item In the $(X,K,A,B)$ scenario, there is a facet for which every Bob setting is indispensable.
 \item A nonsignalling behavior is local if and only if its restriction to every subset of at most $K$ Bob settings is local. The value $K$ is necessary: there exists a nonsignalling behavior with $K$ Bob settings that is nonlocal, although its restriction to any $K-1$ Bob settings is local.
\end{enumerate}
\end{theorem}
\end{minipage}\par

Since quantum behaviors are nonsignalling, every nonlocal quantum behavior has a nonlocal restriction to at most $K$ Bob settings. The example showing that $K$ is necessary is constructed in the nonsignalling set. This leaves open whether, for every choice of $X,A,B\geq2$, there exists a quantum behavior that is nonlocal with all $K$ of Bob's settings but becomes local whenever any one of those settings is removed.

\section{A common distribution over Alice's responses}
\label{sec:extensions}

\subsection{Fixing Alice's observed marginals}

Fix a nonsignalling behavior $p$. Let $Q(\pa)$ be the set of distributions $q=(q_\alpha)_{\alpha\in\Omega}$ over Alice's response strings that reproduce her observed marginals:
\begin{equation}
 Q(\pa)=\left\{q\geq0:\ \sum_\alpha q_\alpha=1,\quad
 \sum_{\alpha:\alpha_x=a}q_\alpha=\pa(a|x)\text{ for every }x,a\right\}.
 \label{eq:Q}
\end{equation}
Here $q_\alpha$ is the probability of the \emph{whole string} $\alpha$. For example, $q_{01}$ for two binary settings is the probability that Alice's predetermined answer is zero at the first setting and one at the second. 

The set $Q(\pa)$ is nonempty: the product distribution
\begin{equation}
 q_\alpha=\prod_{x=1}^X\pa(\alpha_x|x)
 \label{eq:productAlice}
\end{equation}
has the required marginals. An \emph{extension} of Alice's observed marginals is a joint distribution over her predetermined answers $(\alpha_1,\ldots,\alpha_X)$ in which the probability that $\alpha_x=a$ equals $\pa(a|x)$ for every setting $x$ and outcome $a$. The product distribution gives one such extension by treating the answers as independent, although correlated answers can also reproduce the same marginals.

Once Alice's observed marginals are fixed, we ask how many of the $A^X$ probabilities $q_\alpha$ assigned to her predetermined answer strings can still be chosen independently. To bound this number, we retain the normalization equation for $q$ and, for each setting $x$, the marginal equations for outcomes $a=1,\ldots,A-1$. The equation for outcome $a=0$ at each setting follows from these constraints. These $1+X(A-1)$ equations are independent. In the matrix of these equations, columns are indexed by strings $\alpha$. The normalization row has entry $1$ in every column, whereas the row for the marginal $(x,a)$ has entry $\ind{\alpha_x=a}$. Hence a linear dependence among the rows would imply
\begin{equation}
 c_0+\sum_{x=1}^X\sum_{a=1}^{A-1}c_{xa}\ind{\alpha_x=a}=0
 \quad\text{for every }\alpha.
 \label{eq:marginalrowdependence}
\end{equation}
The all-zero string gives $c_0=0$. A string whose only nonzero entry is $\alpha_x=a$ then gives $c_{xa}=0$. This works for every $x,a$, so the rows must be independent. Thus, with $d=X(A-1)$,
\begin{equation}
 r:=\dim Q(\pa)\leq A^X-1-d=K-1.
 \label{eq:Qdim}
\end{equation}
If every $\pa(a|x)$ is positive, the product distribution in \cref{eq:productAlice} is strictly positive. A sufficiently small neighborhood within the equality-constraint space remains nonnegative, so equality holds in \cref{eq:Qdim}. If a marginal is zero, all strings containing that outcome at that setting have zero weight, and the dimension can be smaller.

\subsection{Two binary settings: a tetrahedron cut by two planes}
\label{sec:tetra}

As an illustration, take $X=A=2$. Before fixing the marginals, the four numbers $q_{00},q_{01},q_{10},q_{11}$ form a probability simplex: a tetrahedron. Its vertices $e_{ij}$ put probability one on the string $(i,j)$.

Write $u=\pa(0|1)$, $v=\pa(0|2)$, and $s=q_{00}$. The table of string probabilities is
\begin{equation}
\begin{array}{c|cc|c}
 &\alpha_2=0&\alpha_2=1&\text{row sum}\\\hline
 \alpha_1=0&s&u-s&u\\
 \alpha_1=1&v-s&1-u-v+s&1-u\\\hline
 \text{column sum}&v&1-v&1
\end{array}.
 \label{eq:binarytable}
\end{equation}
Nonnegativity is equivalent to
\begin{equation}
 \max(0,u+v-1)\leq s\leq\min(u,v).
 \label{eq:sinterval}
\end{equation}
The first marginal fixes the plane $q_{00}+q_{01}=u$, and the second fixes $q_{00}+q_{10}=v$. The common intersection of these planes \emph{inside the tetrahedron} is $Q(\pa)$.

For uniform marginals, $u=v=1/2$, the segment has endpoints
\begin{equation}
\begin{aligned}
 q^-&=(0,\tfrac12,\tfrac12,0)=\tfrac12 e_{01}+\tfrac12 e_{10},\\
 q^+&=(\tfrac12,0,0,\tfrac12)=\tfrac12 e_{00}+\tfrac12 e_{11}.
\end{aligned}
 \label{eq:qendpoints}
\end{equation}
These are the extreme points of $Q(\pa)$, although neither is a vertex of the original tetrahedron. An extreme point is always defined relative to the set under consideration. If, for example, $u=0$, \cref{eq:sinterval} forces $s=0$ and the segment reduces to a single point.

\begin{figure}[tbp]
\centering
\begin{subfigure}[t]{.46\textwidth}
\centering
\begin{tikzpicture}[scale=1.05,font=\small]
 \coordinate (a) at (1.7,3.2);\coordinate (b) at (0,0);
 \coordinate (c) at (3.5,0);\coordinate (d) at (1.85,1.15);
 \fill[navy!4] (a)--(b)--(c)--cycle;
 \draw[navy,dashed] (a)--(d)--(b) (d)--(c);
 \draw[navy,thick] (a)--(b)--(c)--cycle;
 \foreach \p/\lab/\pos in {a/00/above,b/01/below left,c/10/below right,d/11/right}
 {\fill[navy] (\p) circle (2pt);\node[\pos] at (\p) {$e_{\lab}$};}
 \node[align=center] at (1.75,-.8) {All distributions over\\the four Alice strings};
\end{tikzpicture}
\end{subfigure}\hfill
\begin{subfigure}[t]{.51\textwidth}
\centering
\begin{tikzpicture}[scale=1.05,font=\small]
 \coordinate (a) at (1.7,3.2);\coordinate (b) at (0,0);
 \coordinate (c) at (3.5,0);\coordinate (d) at (1.85,1.15);
 \fill[navy!4] (a)--(b)--(c)--cycle;
 \draw[navy,dashed] (a)--(d)--(b) (d)--(c);
 \filldraw[orange,fill=orange!25,fill opacity=.65]
 ($(a)!.5!(c)$)--($(a)!.5!(d)$)--($(b)!.5!(d)$)--($(b)!.5!(c)$)--cycle;
 \filldraw[teal,fill=teal!25,fill opacity=.65]
 ($(a)!.5!(b)$)--($(a)!.5!(d)$)--($(c)!.5!(d)$)--($(c)!.5!(b)$)--cycle;
 \draw[navy,thick] (a)--(b)--(c)--cycle;
 \draw[teal,line width=3pt] ($(a)!.5!(d)$)--($(b)!.5!(c)$);
 \fill[teal] ($(a)!.5!(d)$) circle (3pt) node[right=5pt] {$q^+$};
 \fill[teal] ($(b)!.5!(c)$) circle (3pt) node[below] {$q^-$};
 \node[above] at (a) {$e_{00}$};\node[below left] at (b) {$e_{01}$};
 \node[below right] at (c) {$e_{10}$};\node[right] at (d) {$e_{11}$};
 \node[align=center] at (1.75,-.8) {Fixed uniform marginals:\\the segment is $Q(\pa)$};
\end{tikzpicture}
\end{subfigure}
\caption{The extension space for two binary Alice settings. Left: normalization and nonnegativity give a tetrahedron with four deterministic vertices. Right: the orange section imposes $q_{00}+q_{01}=1/2$ and the teal section imposes $q_{00}+q_{10}=1/2$. Their common segment has the two endpoints in \cref{eq:qendpoints}. Fixing marginals changes both the dimension and the extreme points of the feasible set.}
\label{fig:tetra}
\end{figure}
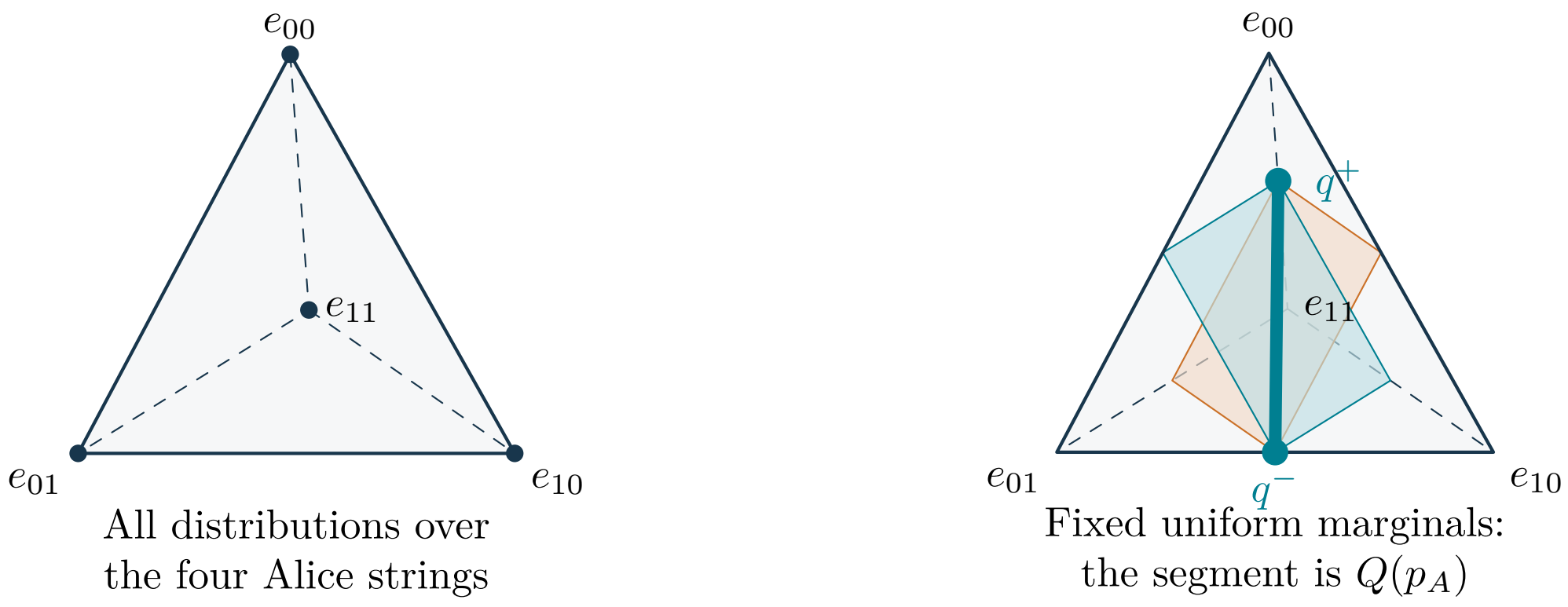

\subsection{Constraints imposed by each Bob setting}

For a chosen $y$, ask which $q\in Q(\pa)$ can reproduce the joint data at that setting. Introduce nonnegative numbers $t^y_{\alpha b}$, interpreted as a joint distribution of Alice's entire response string and Bob's outcome at $y$. They must satisfy
\begin{align}
 \sum_b t^y_{\alpha b}&=q_\alpha
 &&\text{for every }\alpha,\label{eq:witness1}\\
 \sum_{\alpha:\alpha_x=a}t^y_{\alpha b}&=p(a,b|x,y)
 &&\text{for every }a,b,x.\label{eq:witness2}
\end{align}
The first condition preserves the chosen distribution of Alice strings. The second reproduces all the observed pairs involving Bob setting $y$. Define
\begin{equation}
 Q_y(p)=\{q\in Q(\pa):\text{there exists }t^y\geq0
 \text{ satisfying \cref{eq:witness1,eq:witness2}}\}.
 \label{eq:Qy}
\end{equation}

Each $Q_y(p)$ is a convex polytope: it is the projection onto $q$ of the polytope of nonnegative pairs $(q,t^y)$ satisfying \cref{eq:witness1,eq:witness2}.

For nonsignalling data, each individual $Q_y(p)$ is nonempty. Put $w_b=\pb(b|y)$. If $w_b>0$, define
\begin{equation}
 t^y_{\alpha b}=w_b\prod_{x=1}^X\frac{p(\alpha_x,b|x,y)}{w_b}.
 \label{eq:singleBob}
\end{equation}
Each factor is a normalized distribution over $\alpha_x$, since Bob's marginal is independent of $x$. Summing all coordinates except $\alpha_x$ gives \cref{eq:witness2}. If $w_b=0$, nonnegativity forces $p(a,b|x,y)=0$ for every $a,x$, and we set $t^y_{\alpha b}=0$. Taking $q^y_\alpha=\sum_b t^y_{\alpha b}$ then gives a point of $Q_y(p)$.

Different Bob settings can give different $q^y$. The question is whether a \emph{single} $q$ works for all of them.

\subsection{Equivalence with a local model}

The following characterization of locality was established by Pironio~\cite[Lemma~5]{PironioCHSH}, extending Fine's joint-distribution characterization~\cite{Fine}. We state it in our notation and include its proof for completeness.

\begin{proposition}[Common-extension criterion]
\label[proposition]{prop:compat}
A nonsignalling behavior is local if and only if
\begin{equation}
 \bigcap_{y=1}^Y Q_y(p)\neq\varnothing.
 \label{eq:compat}
\end{equation}
\end{proposition}

\begin{proof}
Suppose first that $p$ has the local decomposition \cref{eq:localmixture}. Define
\begin{equation}
 q_\alpha=\sum_\beta\mu_{\alpha\beta},
 \qquad
 t^y_{\alpha b}=\sum_{\beta:\beta_y=b}\mu_{\alpha\beta}.
 \label{eq:localtowitness}
\end{equation}
The first expression is a normalized distribution with Alice's observed marginals. For every $y$,
\begin{align}
 \sum_b t^y_{\alpha b}&=\sum_\beta\mu_{\alpha\beta}=q_\alpha,\\
 \sum_{\alpha:\alpha_x=a}t^y_{\alpha b}
 &=\sum_{\substack{\alpha:\alpha_x=a\\\beta:\beta_y=b}}
 \mu_{\alpha\beta}=p(a,b|x,y).
\end{align}
Thus the same $q$ belongs to every $Q_y(p)$.

Conversely, choose $q$ in the intersection and a witness $t^y$ for each setting. Sample $\alpha$ with probability $q_\alpha$ and let Alice output $a=\alpha_x$. For $q_\alpha>0$, define Bob's conditional response by
\begin{equation}
 R_y(b|\alpha)=\frac{t^y_{\alpha b}}{q_\alpha}.
 \label{eq:R}
\end{equation}
These probabilities sum to one by \cref{eq:witness1}. When $q_\alpha=0$, all $t^y_{\alpha b}$ vanish, and Bob's response at that string can be chosen arbitrarily.

The resulting model gives
\begin{align}
 \sum_\alpha q_\alpha\ind{a=\alpha_x}R_y(b|\alpha)
 &=\sum_{\substack{\alpha:\alpha_x=a\\q_\alpha>0}}t^y_{\alpha b}
 =\sum_{\alpha:\alpha_x=a}t^y_{\alpha b}
 =p(a,b|x,y).
 \label{eq:reconstruct}
\end{align}
Alice's response depends only on $(x,\alpha)$, and Bob's only on $(y,\alpha)$, so this is a local model.

Equivalently, the distribution over deterministic strategies is
\begin{equation}
 \widetilde\mu_{\alpha\beta}=q_\alpha\prod_{y=1}^Y R_y(\beta_y|\alpha).
 \label{eq:deterministicextension}
\end{equation}
\end{proof}

\section{The upper bound from convex intersections}
\label{sec:helly}

\subsection{Helly's theorem}

We use the following finite-dimensional result~\cite{Helly} and include its standard proof based on Radon's affine-dependence argument~\cite{Radon}.

\begin{lemma}[Helly's theorem]
\label[lemma]{lem:helly}
Let $C_1,\ldots,C_m$ be convex sets in an affine space of dimension $r$. If their intersection is empty, some subfamily of at most $r+1$ sets already has empty intersection.
\end{lemma}

\begin{proof}
Choose an inclusion-minimal subfamily with empty intersection and relabel its sets $C_1,\ldots,C_n$. Removing any one set leaves a nonempty intersection. Therefore, for every $i$, there is a point
\begin{equation}
 z_i\in\bigcap_{j\neq i}C_j.
 \label{eq:hellywitnesspoints}
\end{equation}
If $n>r+1$, the points are affinely dependent. There are real numbers $c_i$, not all zero, such that
\begin{equation}
 \sum_i c_i z_i=0,\qquad \sum_i c_i=0.
 \label{eq:hellyaffinedependence}
\end{equation}
The coefficients include both signs. Let $t=\sum_{c_i>0}c_i=-\sum_{c_i<0}c_i>0$. The two convex combinations
\begin{equation}
 z=\sum_{c_i>0}\frac{c_i}{t}z_i
 =\sum_{c_i<0}\frac{-c_i}{t}z_i
 \label{eq:radon}
\end{equation}
are equal. For any fixed $k$, at least one of these combinations omits $z_k$. Every point in that combination belongs to $C_k$, since $z_i\in C_k$ whenever $i\neq k$. Convexity implies $z\in C_k$. This holds for every $k$, contradicting the empty intersection. Hence $n\leq r+1$.
\end{proof}

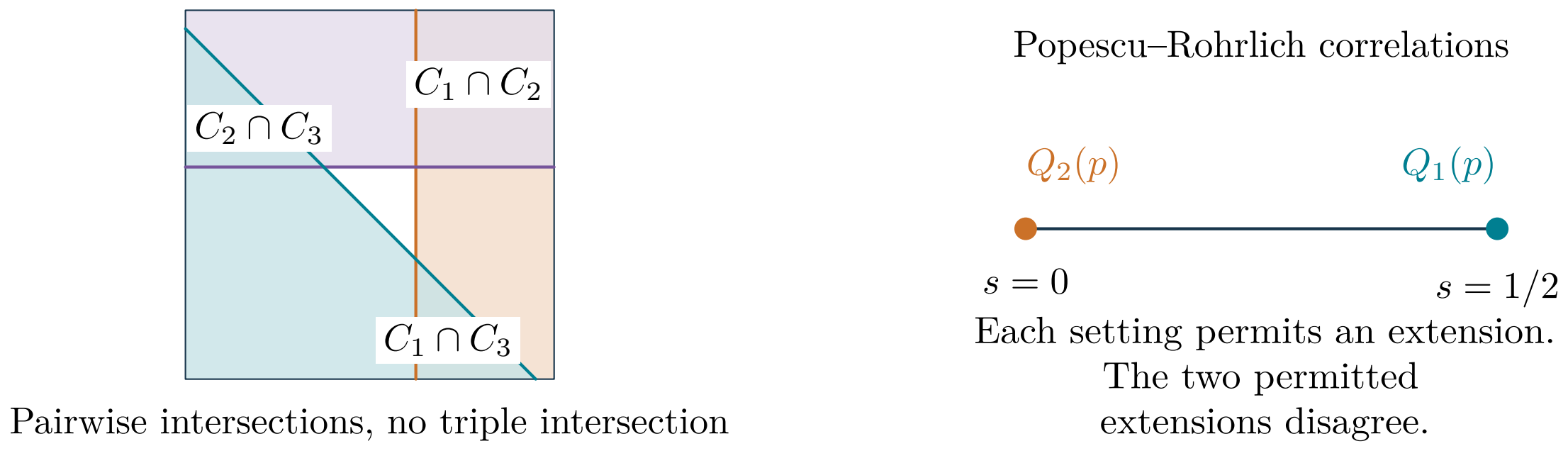
\begin{figure}[tbp]
\centering
\begin{subfigure}[t]{.49\textwidth}
\centering
\begin{tikzpicture}[scale=.86,font=\small]
 \fill[orange!20] (2.5,0) rectangle (4,4);
 \fill[purple!20,opacity=.8] (0,2.3) rectangle (4,4);
 \fill[teal!22,opacity=.8] (0,0)--(3.8,0)--(0,3.8)--cycle;
 \draw[navy] (0,0) rectangle (4,4);
 \draw[orange,thick] (2.5,0)--(2.5,4);
 \draw[purple,thick] (0,2.3)--(4,2.3);
 \draw[teal,thick] (0,3.8)--(3.8,0);
 \node[fill=white,inner sep=2pt] at (3.18,3.2) {$C_1\cap C_2$};
 \node[fill=white,inner sep=2pt] at (.8,2.72) {$C_2\cap C_3$};
 \node[fill=white,inner sep=2pt] at (2.85,.42) {$C_1\cap C_3$};
 \node[below] at (2,-.15) {Pairwise intersections, no triple intersection};
\end{tikzpicture}
\end{subfigure}\hfill
\begin{subfigure}[t]{.47\textwidth}
\centering
\begin{tikzpicture}[font=\small]
 \draw[navy,thick] (0,2)--(4.4,2);
 \fill[orange] (0,2) circle (3pt);
 \fill[teal] (4.4,2) circle (3pt);
 \node[below=7pt] at (0,2) {$s=0$};
 \node[below=7pt] at (4.4,2) {$s=1/2$};
 \node[above=8pt,orange] at (.45,2) {$Q_2(p)$};
 \node[above=8pt,teal] at (3.95,2) {$Q_1(p)$};
 \node[above] at (2.2,3.4) {Popescu--Rohrlich correlations};
 \node[align=center,text width=5.7cm] at (2.2,.6)
 {Each setting permits an extension.\\The two permitted extensions disagree.};
\end{tikzpicture}
\end{subfigure}
\caption{Left: an illustration of Helly's theorem in a plane. Inside $[0,4]^2$, take $C_1=\{u\geq2.5\}$, $C_2=\{v\geq2.3\}$, and $C_3=\{u+v\leq3.8\}$. Each pair intersects, but all three cannot, since $2.5+2.3>3.8$. Right: for the Popescu--Rohrlich example in \cref{sec:PR}, the extension space is a line segment and the two Bob settings select opposite endpoints. The left panel illustrates the convex theorem; it is not asserted to be an extension space of a particular Bell behavior.}
\label{fig:helly}
\end{figure}

The number $r+1$ matters. In a plane, three convex sets can intersect pairwise and still have no common point. \Cref{fig:helly} shows such an example. For intervals on a line, in contrast, pairwise intersection is sufficient.

\subsection{Applying the theorem to restricted Bell data}

If $S$ is a subset of Bob settings, restricting to $S$ preserves Alice's marginals. Moreover, $Q_y(p)$ depends only on the data at $y$. Applying \cref{prop:compat} to the restricted behavior gives
\begin{equation}
 \pi_S(p)\in\Lloc^{A,B}_{X,|S|}
 \quad\Longleftrightarrow\quad
 \bigcap_{y\in S}Q_y(p)\neq\varnothing.
 \label{eq:restrictedcompat}
\end{equation}
All these convex sets lie in the common affine space $\aff Q(\pa)$, whose dimension satisfies $r\leq K-1$.

Suppose every restriction to at most $K$ settings is local. By \cref{eq:restrictedcompat}, every subfamily of at most $r+1\leq K$ compatibility sets has a common point. Helly's theorem implies that all $Q_y(p)$ have a common point. \Cref{prop:compat} then gives a local model for the full behavior. The converse is immediate: discarding settings from a local model leaves a local model. We have proved
\begin{equation}
 p\text{ is local}\quad\Longleftrightarrow\quad
 \pi_S(p)\text{ is local for all }S\text{ with }1\leq|S|\leq K.
 \label{eq:localitycutoff}
\end{equation}
Subsets always lie inside the actual set of $Y$ settings; when $Y\leq K$, the full set is included.

Bob's number of outcomes affects the witness arrays $t^y$ and may change the shape of $Q_y(p)$. It does not change the variable $q$ or the dimension bound \cref{eq:Qdim}. This is why the upper cutoff is independent of $B$.

\subsection{An explicit incompatible pair: the PR correlations}
\label{sec:PR}

Take settings $x,y\in\{1,2\}$ and outcomes $a,b\in\{0,1\}$. The Popescu--Rohrlich (PR) \mbox{correlations}~\cite{PopescuRohrlich} are given by
\begin{equation}
 p(a,b|x,y)=
 \begin{cases}
 1/2,&a\oplus b=(x-1)(y-1),\\
 0,&\text{otherwise}.
 \end{cases}
 \label{eq:PR}
\end{equation}
An Alice response string is $\alpha=(\alpha_1,\alpha_2)$, where $\alpha_x$ is her predetermined output for setting $x$, and $q_{ij}$ is the probability that $(\alpha_1,\alpha_2)=(i,j)$. Alice's marginals are uniform, so her possible extensions are
\begin{equation}
 q=(q_{00},q_{01},q_{10},q_{11})=(s,\tfrac12-s,\tfrac12-s,s),\qquad 0\leq s\leq\tfrac12.
\end{equation}

At $y=1$, Bob's answer must agree with Alice's answer at both settings. Any compatible extension therefore has $\alpha_1=\alpha_2$ with probability one, forcing $s=1/2$. This endpoint is compatible by choosing Bob's output equal to their common value. Thus $Q_1(p)=\{q^+\}$.

At $y=2$, Bob must agree with $\alpha_1$ and disagree with $\alpha_2$. Hence $\alpha_1\neq\alpha_2$ with probability one, forcing $s=0$, and this endpoint is compatible by setting $b=\alpha_1$. Thus $Q_2(p)=\{q^-\}$. Each setting separately admits a local model, but no common Alice distribution works for both.

\section{Why every facet involves at most \texorpdfstring{$K$}{K} Bob settings}
\label{sec:facetupper}

We have proved that a nonsignalling behavior is local if and only
if its restriction to every subset of at most $K$ Bob settings is
local. We now show that every tight Bell inequality can therefore
be written using at most $K$ Bob settings.

For each nonempty set $S\subseteq\{1,\ldots,Y\}$ of at most $K$ Bob settings,
consider all facet inequalities of the restricted local polytope
$\pi_S(\Lloc^{A,B}_{X,Y})=\Lloc^{A,B}_{X,|S|}$. Write them as
\begin{equation}
 f_{S,\nu}(u)\geq0,\qquad \nu=1,\ldots,N_S,
 \label{eq:restrictedfacetlist}
\end{equation}
where $\nu$ labels individual facets, including those related by relabellings.
The list includes all positivity facets $p(a,b|x,y)\geq0$ with $y\in S$, including when $S$ contains a single Bob setting~\cite[Theorem~2]{Pironio}.
We lift each inequality to the full scenario by keeping its coefficients on $S$ and
assigning zero coefficients to all terms involving Bob settings $y\notin S$. Evaluating the lifted expression on $p$
then gives the same value as evaluating the original expression on $\pi_S(p)$.
The locality criterion therefore gives
\begin{equation}
 \Lloc^{A,B}_{X,Y}
 =
 \left\{p\in\Naff^{A,B}_{X,Y}:
 f_{S,\nu}\bigl(\pi_S(p)\bigr)\geq0
 \text{ for every }S,\nu\right\}.
 \label{eq:finiteH}
\end{equation}
Every local behavior satisfies these inequalities. Conversely, suppose that a point
$p\in\Naff^{A,B}_{X,Y}$ satisfies them all. For every nonempty $S$ with $|S|\leq K$, the restriction $\pi_S(p)$ satisfies every facet inequality of the restricted local polytope and is therefore local. In particular, for $S=\{y\}$, we keep all of Alice's settings and every outcome for both parties, but only Bob's setting $y$. Locality of this restriction implies $p(a,b|x,y)\geq0$ for every $a,b,x$. Taking each $y$ in turn shows that every entry of $p$ is nonnegative. Thus $p$ is a nonsignalling behavior and all the restrictions in \cref{eq:localitycutoff} are local, so $p$ itself is local.

Choose a facet $\mathcal T$ of the full local polytope. Let $v_1,\ldots,v_m$ be
all the deterministic vertices that saturate the inequality defining $\mathcal T$,
and let $p^\circ=m^{-1}\sum_{i=1}^m v_i$ be their average.

Suppose every inequality in \eqref{eq:finiteH} were strictly satisfied at $p^\circ$.
Since the family of inequalities in \eqref{eq:finiteH} is finite,
all of them would remain strictly satisfied throughout a
sufficiently small neighborhood of $p^\circ$ in the nonsignalling
affine space. By \eqref{eq:finiteH}, that entire
neighborhood would be local, making $p^\circ$ an interior point of the local polytope.
But $p^\circ$ lies on the facet $\mathcal T$, so it must saturate at least one
inequality in \eqref{eq:finiteH}.

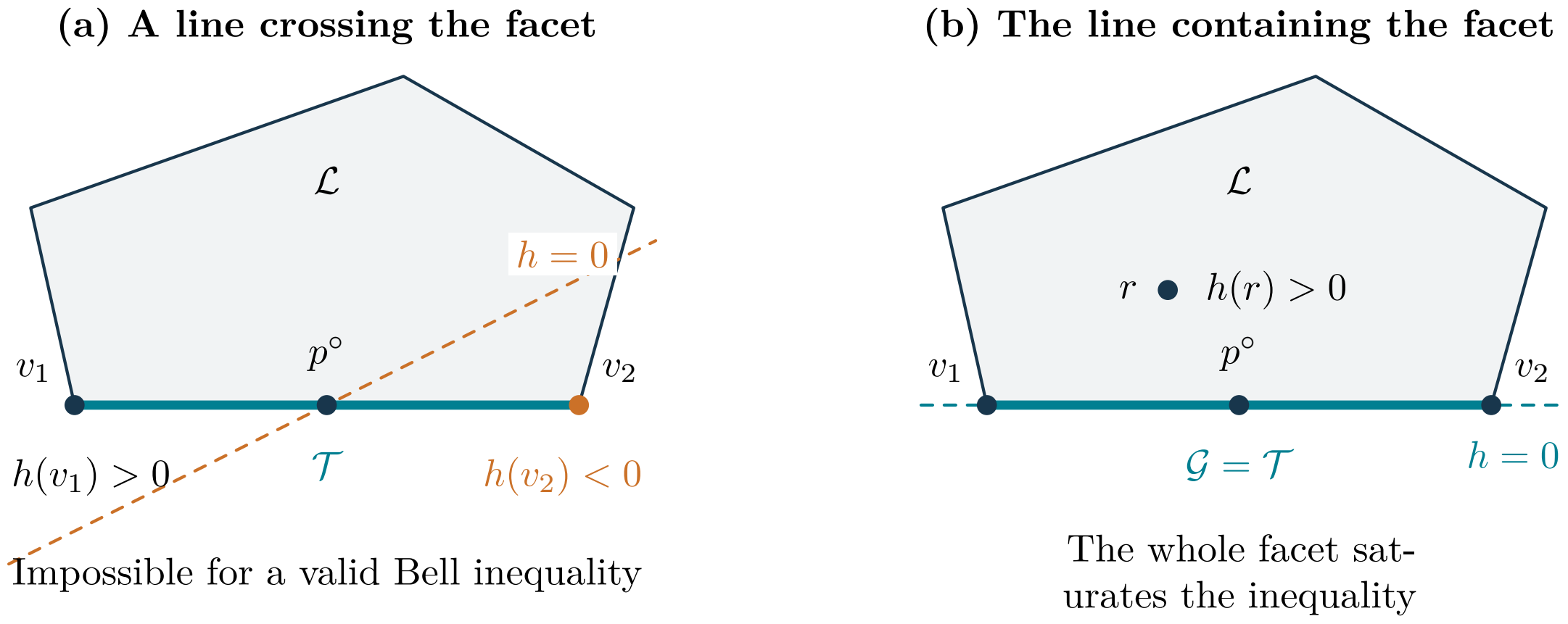
\begin{figure}[tbp]
\centering

\begin{minipage}[t]{0.48\textwidth}
\centering
\begin{tikzpicture}[font=\small]
 \path[use as bounding box]
   (-0.95,-1.75) rectangle (5.55,3.65);

 \filldraw[draw=navy,fill=navy!6,thick]
   (0,0)--(4.6,0)--(5.1,1.8)--(3,3)--(-0.4,1.8)--cycle;

 \node[font=\bfseries\small] at (2.3,3.45)
   {(a) A line crossing the facet};

 \node at (2.3,2.05) {$\Lloc$};

 \draw[teal,line width=2.4pt] (0,0)--(4.6,0);
 \node[teal] at (2.3,-0.55) {$\mathcal T$};

 \draw[orange,dashed,thick]
   (-0.6,-1.45)--(5.3,1.5);
 \node[orange,fill=white,inner sep=2pt]
   at (4.45,1.38) {$h=0$};

 \fill[navy] (0,0) circle (2.6pt);
 \fill[orange] (4.6,0) circle (2.6pt);
 \fill[navy] (2.3,0) circle (2.6pt);

 \node[above left=3pt] at (0,0) {$v_1$};
 \node[above right=3pt] at (4.6,0) {$v_2$};
 \node[above=5pt] at (2.3,0) {$p^\circ$};

 \node at (0.15,-0.65) {$h(v_1)>0$};
 \node[orange] at (4.45,-0.65) {$h(v_2)<0$};

 \node[align=center,text width=6cm] at (2.3,-1.55)
   {Impossible for a valid Bell inequality};
\end{tikzpicture}
\end{minipage}
\hfill
\begin{minipage}[t]{0.48\textwidth}
\centering
\begin{tikzpicture}[font=\small]
 \path[use as bounding box]
   (-0.95,-1.75) rectangle (5.55,3.65);

 \filldraw[draw=navy,fill=navy!6,thick]
   (0,0)--(4.6,0)--(5.1,1.8)--(3,3)--(-0.4,1.8)--cycle;

 \node[font=\bfseries\small] at (2.3,3.45)
   {(b) The line containing the facet};

 \node at (2.3,2.05) {$\Lloc$};

 \draw[teal,dashed,thick] (-0.6,0)--(5.3,0);
 \draw[teal,line width=2.4pt] (0,0)--(4.6,0);

 \node[teal] at (2.3,-0.55)
   {$\mathcal G=\mathcal T$};
 \node[teal] at (4.8,-0.45) {$h=0$};

 \fill[navy] (0,0) circle (2.6pt);
 \fill[navy] (4.6,0) circle (2.6pt);
 \fill[navy] (2.3,0) circle (2.6pt);

 \node[above left=3pt] at (0,0) {$v_1$};
 \node[above right=3pt] at (4.6,0) {$v_2$};
 \node[above=5pt] at (2.3,0) {$p^\circ$};

 \fill[navy] (1.65,1.05) circle (2.6pt);
 \node[left=4pt] at (1.65,1.05) {$r$};
 \node[right=6pt] at (1.65,1.05) {$h(r)>0$};

 \node[align=center,text width=6cm] at (2.3,-1.55)
   {The whole facet saturates the inequality};
\end{tikzpicture}
\end{minipage}

\caption{Why a lifted inequality must define the whole facet.
In this two-dimensional illustration, the local polytope is
represented by a polygon, the facet $\mathcal T$ is an edge
with endpoints $v_1,v_2$, and $p^\circ=(v_1+v_2)/2$.
(a) A line $h=0$ passing through $p^\circ$ but crossing the edge
puts one endpoint on the side $h<0$. This is impossible for an
inequality valid on all local behaviors.
(b) The valid inequality must therefore be saturated on the
entire edge, giving $\mathcal G=\mathcal T$.
The local point $r$ with $h(r)>0$ shows that this saturating
face is smaller than the whole polytope.
In higher dimensions, the proof uses the same argument with
the average of all vertices of the facet.}
\label{fig:facetupper}
\end{figure}

Choose one such inequality, indexed by $(S,\nu)$, and write its lifted expression as
$h(p)=f_{S,\nu}(\pi_S(p))$. Saturation gives $h(p^\circ)=0$. Since $h$ is affine,
\begin{equation}
 h(p^\circ)=\frac{1}{m}\sum_{i=1}^m h(v_i).
 \label{eq:facetaverage}
\end{equation}
Each $h(v_i)$ is nonnegative because $v_i$ is local. Their average is zero, so
every $v_i$ also saturates $h\geq0$. Hence the face
$\mathcal G=\{p\in\Lloc^{A,B}_{X,Y}:h(p)=0\}$ contains $\mathcal T$;
see \cref{fig:facetupper}.

The face $\mathcal G$ is proper. Indeed, some restricted local behavior $u$ satisfies $f_{S,\nu}(u)>0$. Extending its local model by fixing the omitted Bob outputs gives a full local behavior with $h>0$. Since a facet is a maximal proper face, $\mathcal G=\mathcal T$. Thus the chosen facet is the input lifting of a facet on $S$, with $|S|\leq K$.

\section{A criterion for tightness}
\label{sec:criterion}

We first construct the inequalities for binary Bob outcomes, $b_y\in\{0,1\}$. Our starting point is to give every deterministic Alice string at least one Bob response that saturates the inequality. We write the inequality as $F(p)\geq0$, so each deterministic behavior that saturates it satisfies $F(p)=0$. We then show that any affine Bell expression that vanishes on all these behaviors is proportional to $F$ on the normalized nonsignalling affine space. This proves that $F(p)\geq0$ defines a facet of the local polytope. The extension to arbitrary $B$ is given in \cref{sec:Boboutcomes}.

\subsection{Decomposing a Bell expression by Bob's settings}

Let $\U$ be the space of functions of Alice's string that are sums of single-setting contributions:
\begin{equation}
 \U=\left\{h:\Omega\longrightarrow\R:
 h(\alpha)=h_0+\sum_{x=1}^X u_x(\alpha_x)\right\}.
 \label{eq:additive}
\end{equation}
We call such functions \emph{additive}. For example, $\ind{\alpha_1=1}-2\ind{\alpha_2=0}$ is additive, whereas $\ind{\alpha_1=1}\ind{\alpha_2=1}$ generally is not. The latter depends jointly on two different answers of Alice, and is not an observable in a standard Bell experiment.

We may always impose $u_x(0)=0$ by absorbing constants into $h_0$. This gives a unique representation and
\begin{equation}
 \dim\U=1+X(A-1)=d+1.
 \label{eq:additivedimension}
\end{equation}
Let $F$ be any affine Bell expression. We write $F(\alpha,\bm b)$ for its value on the deterministic behavior specified by Alice's answer string $\alpha$ and Bob's binary answer string $\bm b$. Using the independent coordinates in \cref{eq:coords}, we can write
\begin{equation}
 F(\alpha,\bm b)=h(\alpha)+\sum_{y=1}^Y b_y h_y(\alpha),
 \qquad h,h_y\in\U.
 \label{eq:Fbinary}
\end{equation}
Here $h$ and each $h_y$ are additive as in \eqref{eq:additive}. The term $h$ collects the constant and Alice's marginals; $h_y$ collects Bob's marginal at $y$ and the joint terms involving that setting.

This decomposition is unique as a collection of functions. Setting all $b_y=0$ determines $h$. Setting only $b_y=1$ then determines $h_y$ by subtraction. Thus we can reason about the functions $h_y$ without ambiguity from normalization or nonsignalling identities.

\subsection{Attaining the local bound for every deterministic Alice strategy}

We first consider binary outputs for Bob, so a deterministic
strategy specifies an answer $b_y\in\{0,1\}$ for each setting $y$.
We write Bell inequalities as $F(p)\geq0$, with the local bound
included in the expression: if the inequality is
$\mathcal I(p)\leq L$, then $F(p)=L-\mathcal I(p)$.
A local behavior saturates the inequality when $F(p)=0$.

For each Bob setting $y$, we choose a real-valued additive
function of Alice's response string,
\begin{equation}
 \ell_y(\alpha)=c_y+\sum_{x=1}^X u_{xy}(\alpha_x).
 \label{eq:settingadditivefunction}
\end{equation}
This function collects the coefficients associated with Bob's
setting $y$. Its contribution to the Bell expression will be
$b_y\ell_y(\alpha)$. For a fixed Alice strategy $\alpha$, Bob
therefore contributes either zero, by answering $b_y=0$, or
$\ell_y(\alpha)$, by answering $b_y=1$.

The values of $\ell_y$ may be positive, negative, or zero.
For example, with two binary Alice settings, the choice
$\ell_y(\alpha)=\alpha_1-\alpha_2$ gives
\begin{equation}
\begin{array}{c|rrrr}
 (\alpha_1,\alpha_2)&(0,0)&(0,1)&(1,0)&(1,1)\\ \hline
 \ell_y(\alpha)&0&-1&1&0
\end{array}.
 \label{eq:binarycoefficientexample}
\end{equation}
For the string $(0,1)$, Bob lowers the expression by answering
$1$; for $(1,0)$, he minimizes his contribution by answering
$0$. For the other two strings, his answer makes no difference.
Other choices of coefficients can give other real values.
We assume that no $\ell_y$ is identically zero, although it
may vanish on some Alice strings.

For each fixed $\alpha$, the smallest contribution Bob can
obtain at setting $y$ is
\begin{equation}
 m_y(\alpha)
 :=\min_{b_y\in\{0,1\}}b_y\ell_y(\alpha)
 =\min\{0,\ell_y(\alpha)\}.
 \label{eq:minima}
\end{equation}
The minimization is over Bob's answer, with Alice's entire
response string held fixed. Since Bob can assign an answer
to each setting independently, his minimum total contribution
is $\sum_y m_y(\alpha)$.

To obtain a saturating response for every Alice string, we choose the Alice-only term to cancel this minimum:
\begin{equation}
 g(\alpha):=-\sum_y m_y(\alpha).
 \label{eq:gadditive}
\end{equation}
We require that $g$ be additive, that is, $g\in\U$.
This is a condition on our choice of the functions $\ell_y$;
it does not hold for arbitrary choices. In \eqref{eq:Fbinary}, set
$h=g$ and $h_y=\ell_y$. We claim that
\begin{equation}
 F(\alpha,\bm b)
 =g(\alpha)+\sum_y b_y\ell_y(\alpha)\geq0
 \label{eq:generalF}
\end{equation}
is a valid Bell inequality with local minimum zero.

To prove nonnegativity, use the definition of $g$ to write
\begin{equation}
 F(\alpha,\bm b)
 =\sum_y\bigl[b_y\ell_y(\alpha)-m_y(\alpha)\bigr].
 \label{eq:nonnegativeterms}
\end{equation}
Each term is nonnegative because $m_y(\alpha)$ is the minimum
of $b_y\ell_y(\alpha)$. Thus $F\geq0$ for every deterministic
strategy. Moreover, choosing a minimizing answer at every
Bob setting gives
\begin{equation}
 \min_{\bm b}F(\alpha,\bm b)
 =g(\alpha)+\sum_y m_y(\alpha)=0.
 \label{eq:Fminimum}
\end{equation}
Every Alice string therefore admits at least one saturating
Bob response.

We now check that $F$ can indeed be written using the observed
Bell probabilities. Since $g$ and the functions $\ell_y$ are
additive, write
\begin{equation}
 g(\alpha)=g_0+\sum_x g_x(\alpha_x),
 \qquad
 \ell_y(\alpha)=c_y+\sum_x u_{xy}(\alpha_x).
 \label{eq:additivecoefficientforms}
\end{equation}
On deterministic strategies, the corresponding expression is
\begin{equation}
 F(\alpha,\bm b)
 =g_0+\sum_x g_x(\alpha_x)
 +\sum_y c_y b_y
 +\sum_{x,y}b_yu_{xy}(\alpha_x).
 \label{eq:deterministicexpanded}
\end{equation}
Its expression in observed probabilities is therefore
\begin{equation}
\begin{aligned}
 F(p)={}&g_0
 +\sum_{x=1}^X\sum_{a=0}^{A-1}
       g_x(a)\,\pa(a|x)
 +\sum_{y=1}^Y c_y\,\pb(1|y)
 \\
 &+\sum_{x=1}^X\sum_{y=1}^Y\sum_{a=0}^{A-1}
       u_{xy}(a)\,p(a,1|x,y).
\end{aligned}
 \label{eq:generalFobserved}
\end{equation}
This contains only Alice's marginals, Bob's marginals, and
Alice--Bob joint probabilities. Since it is affine in $p$,
its nonnegativity on deterministic strategies implies
nonnegativity on every local mixture. Together with
\cref{eq:Fminimum}, this establishes the local minimum zero.

The additivity requirement is essential here. Taking the
minimum in \cref{eq:minima} can produce dependence on several
Alice answers jointly, even though $\ell_y$ is additive.
Such dependence would generally require probabilities for
the outcomes of different Alice settings in the same run,
which are not part of the observed Bell data. The individual
$m_y$ may contain these terms, but their sum must cancel them
so that $g$ is additive.

Finally, the nonnegative-term decomposition of $F$ identifies
all saturating deterministic strategies. Equality holds if
and only if Bob minimizes every term:
\begin{equation}
 b_y=
 \begin{cases}
  1,&\ell_y(\alpha)<0,\\
  0,&\ell_y(\alpha)>0,\\
  0\ \text{or}\ 1,&\ell_y(\alpha)=0.
 \end{cases}
 \label{eq:saturatingrule}
\end{equation}
We call $\ell_y(\alpha)=0$ a \emph{tie}: both outputs give the same Bell value at this Alice string.
Keeping minimizing answers at all other settings, switching
only $b_y$ therefore gives two saturating deterministic
strategies. These pairs will provide the equations used
below to prove that the inequality defines a facet.

\subsection{Using saturating strategies to prove that a Bell inequality defines a facet}

Let $\mathcal V_0$ be the set of saturating deterministic vertices. Suppose their affine span has dimension $s$. In a $D$-dimensional affine space, the vector space of affine functions has dimension $D+1$. Requiring such a function to vanish on an $s$-dimensional affine subspace imposes $s+1$ independent linear conditions. Thus the space of affine functions vanishing on $\mathcal V_0$ has dimension
\begin{equation}
 (D+1)-(s+1)=D-s.
 \label{eq:annihilatordim}
\end{equation}
This can also be seen by using coordinates $(1,p_1,\ldots,p_D)$: $s+1$ affinely independent points become $s+1$ linearly independent rows.

Consequently, if every affine expression vanishing on $\mathcal V_0$ is proportional to the nonzero expression defining the inequality, then $D-s=1$. The saturating face has dimension $D-1$ and the inequality is tight. There is no need to identify explicitly $D$ affinely independent saturating vertices.

The following criterion has two steps. First, the strings at which Bob can switch his output without changing the Bell value fix the coefficient function for each setting, up to a scalar. Second, the other saturation equations force these scalars to agree across settings.

\begin{proposition}[A criterion for tightness]
\label[proposition]{prop:facetcriterion}
Let $\ell_1,\ldots,\ell_Y\in\U$ be nonzero functions, define $m_y$ and $g$ by \cref{eq:minima,eq:gadditive}, and assume that $g\in\U$. Suppose the following two properties are satisfied.
\begin{enumerate}[label=(\roman*)]
 \item For each $y$, if an additive function $h$ vanishes whenever $\ell_y(\alpha)=0$, then $h=\lambda\ell_y$ for some scalar $\lambda$.
 \item If $\sum_y\lambda_y m_y$ is additive, then all coefficients $\lambda_y$ are equal.
\end{enumerate}
Then \cref{eq:generalF} defines a facet. Moreover, every setting with $\ell_y\neq0$ is indispensable.
\end{proposition}

\begin{proof}
For this proof, let $\widetilde F$ denote the Bell expression in \eqref{eq:generalF} for the chosen functions $\ell_1,\ldots,\ell_Y$. Let $F$ be any affine Bell expression of the form \eqref{eq:Fbinary} that vanishes on every deterministic behavior saturating $\widetilde F\geq0$. We will show that $F=\lambda\widetilde F$ for some scalar $\lambda$.

Fix a setting $y$ and an Alice string with $\ell_y(\alpha)=0$. Choose minimizing outputs at all other settings. Both choices $b_y=0$ and $b_y=1$ give saturating strategies. Subtracting their two equations $F=0$ gives
\begin{equation}
 h_y(\alpha)=0\qquad\text{whenever }\ell_y(\alpha)=0.
\end{equation}
Property (i) therefore gives, separately for each setting,
\begin{equation}
 h_y=\lambda_y\ell_y.
 \label{eq:hylambda}
\end{equation}
Now choose minimizing responses at every setting, for any fixed $\alpha$. The equation $F=0$ becomes
\begin{equation}
 h(\alpha)+\sum_y\lambda_y m_y(\alpha)=0
 \qquad\text{for every }\alpha.
 \label{eq:hminrelation}
\end{equation}
Since $h$ is additive, property (ii) applies and all $\lambda_y$ equal a common $\lambda$. Using \cref{eq:gadditive} gives $h=\lambda g$. Together with \cref{eq:hylambda}, this gives $F=\lambda\widetilde F$ on every deterministic behavior. The deterministic behaviors affinely span the normalized nonsignalling affine space, so the same equality holds throughout that space. The dimension argument in \cref{eq:annihilatordim} establishes tightness.

Finally, if $\ell_y\neq0$, choose an $\alpha$ for which $\ell_y(\alpha)\neq0$ and compare two deterministic strategies differing only at $b_y$. Changing $b_y$ from zero to one changes $\widetilde F$ by $\ell_y(\alpha)$. After discarding setting $y$, these two behaviors are identical. Therefore no affine functional of the remaining data can equal $\widetilde F$. This setting is indispensable, including after any use of affine identities.
\end{proof}

For $Y\geq2$, property (ii) also implies that every $\ell_y$ takes a strictly negative value at some Alice string. Indeed, if $m_{y_0}$ were identically zero, choosing $\lambda_{y_0}=1$ and $\lambda_y=0$ for $y\neq y_0$ would give $\sum_y\lambda_y m_y=0\in\U$, although the coefficients are not all equal. We will use this consequence when extending the constructions to more Bob outcomes.

\section{A construction for three or more Alice settings}
\label{sec:lines}

\subsection{Partitioning Alice's response strings}

Assume $X\geq3$. Indices are cyclic, so the predecessor of $1$ is $X$. For each $i=1,\ldots,X$, define
\begin{equation}
 L_i=\left\{\alpha\in\Omega:
 \alpha_{i-1}=1,\quad
 \alpha_j=0\text{ for }j\notin\{i-1,i\}\right\}.
 \label{eq:lines}
\end{equation}
Coordinate $i$ is free and takes all $A$ values. Every other coordinate is fixed. We call $L_i$ a coordinate line because its points differ only in one coordinate.

These $X$ lines are pairwise disjoint. To check this, let $i\neq j$.
If $j\neq i-1$, coordinate $i-1$ equals one on $L_i$. On $L_j$ it is neither the free coordinate $j$ nor the coordinate $j-1$ fixed to one, so it is zero. If $j=i-1$, use coordinate $j-1$ instead. It is one on $L_j$, while on $L_i$ it is fixed to zero: for $X\geq3$, $j-1$ differs from both $i$ and $i-1$. In either case no string belongs to both lines.

Let $\Ccal$ be the partition of $\Omega$ consisting of these lines and every uncovered string as a singleton. There are $X$ line cells containing $AX$ strings in total. The number of cells is therefore
\begin{equation}
 |\Ccal|=X+(A^X-AX)=A^X-X(A-1)=K.
 \label{eq:cellcount}
\end{equation}
Each cell will label one Bob setting.

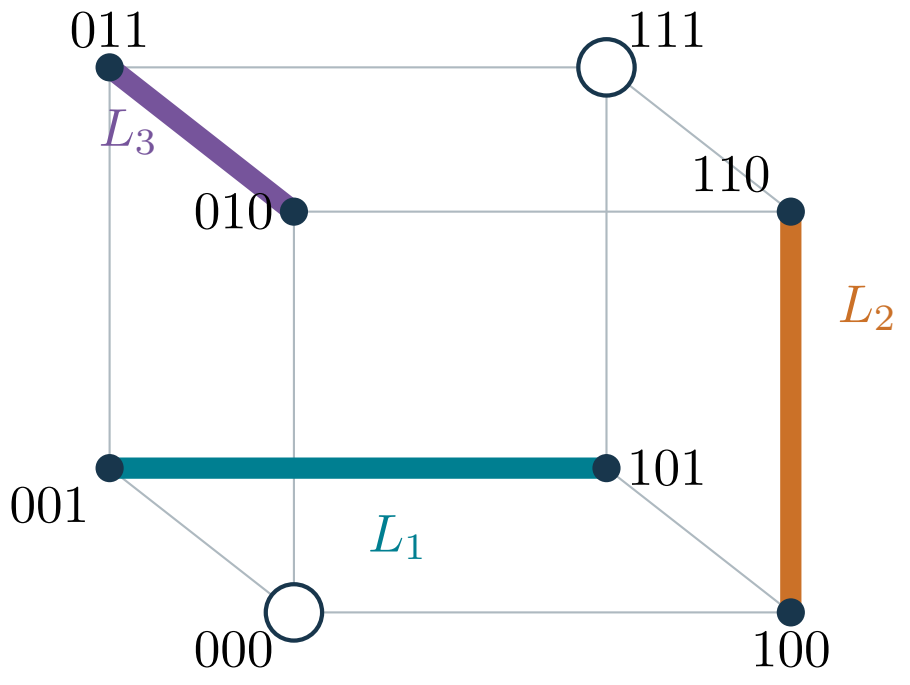
\begin{figure}[tbp]
\centering
\begin{tikzpicture}[font=\small,scale=1.06]
 \coordinate (000) at (1.15,0); \coordinate (100) at (4.25,0);
 \coordinate (010) at (1.15,2.5); \coordinate (110) at (4.25,2.5);
 \coordinate (001) at (0,.9); \coordinate (101) at (3.1,.9);
 \coordinate (011) at (0,3.4); \coordinate (111) at (3.1,3.4);
 \draw[navy!35] (000)--(100)--(110)--(010)--cycle;
 \draw[navy!35] (001)--(101)--(111)--(011)--cycle;
 \draw[navy!35] (000)--(001) (100)--(101) (010)--(011) (110)--(111);
 \draw[teal,line width=4pt] (001)--(101);
 \draw[orange,line width=4pt] (100)--(110);
 \draw[purple,line width=4pt] (010)--(011);
 \foreach \p in {100,110,001,101,010,011}{\fill[navy] (\p) circle (2.5pt);}
 \foreach \p in {000,111}{\filldraw[navy,fill=white,thick] (\p) circle (5pt);}
 \node[below left] at (000) {$000$};\node[below] at (100) {$100$};
 \node[left] at (010) {$010$};\node[above left] at (110) {$110$};
 \node[below left] at (001) {$001$};\node[right] at (101) {$101$};
 \node[above] at (011) {$011$};\node[above right] at (111) {$111$};
 \node[teal,below=5pt] at (1.8,.9) {$L_1$};
 \node[orange,right=5pt] at (4.25,1.9) {$L_2$};
 \node[purple,left=5pt] at (.6,3) {$L_3$};
 \node[align=left,text width=5.5cm] at (8.3,1.8)
 {$L_1$: vary the first answer.\\[4pt]
 $L_2$: vary the second answer.\\[4pt]
 $L_3$: vary the third answer.\\[8pt]
 Three lines and two singletons\\give five Bob settings.};
\end{tikzpicture}
\caption{The lower-bound construction for three binary Alice settings. The eight cube vertices are Alice's deterministic response strings. The three colored edges are disjoint, with one edge parallel to each coordinate direction. The two circled vertices are singleton cells. Every string belongs to exactly one cell, and the five cells label Bob's settings.}
\label{fig:cube}
\end{figure}

For $X=3$ and $A=2$, the construction is particularly simple:
\begin{equation}
 \begin{aligned}
 L_1&=\{001,101\},&
 L_2&=\{100,110\},&
 L_3&=\{010,011\},\\
 C_4&=\{000\},&
 C_5&=\{111\}.
 \end{aligned}
 \label{eq:cubecells}
\end{equation}
These are three disjoint edges of a cube and the two remaining vertices; see \cref{fig:cube}.

\subsection{From a cell to a Bell coefficient}

Each cell $C$ fixes a nonempty set of coordinates $J_C$ to values $c_j$. The prescribed values $c_j$ depend on $C$; we suppress this dependence in the notation. For a line, $|J_C|=X-1$; for a singleton, $|J_C|=X$. Define
\begin{equation}
 \ell_C(\alpha)=\sum_{j\in J_C}\ind{\alpha_j\neq c_j}-1.
 \label{eq:cellfunction}
\end{equation}
This is an additive function. If the string belongs to $C$, there are no mismatches and $\ell_C=-1$. With exactly one fixed-coordinate mismatch, $\ell_C=0$. With two or more, $\ell_C\geq1$. Therefore
\begin{equation}
 m_C(\alpha)=\min\{0,\ell_C(\alpha)\}=-\ind{\alpha\in C}.
 \label{eq:cellminimum}
\end{equation}
Because the cells partition $\Omega$, exactly one indicator on the right equals one for every $\alpha$. Hence
\begin{equation}
 \sum_{C\in\Ccal}m_C(\alpha)=-1.
 \label{eq:partitionsum}
\end{equation}
The additive function $g$ in \cref{eq:gadditive} is simply the constant one, and the inequality is
\begin{equation}
 F(\alpha,\bm b)=1+\sum_{C\in\Ccal}b_C\ell_C(\alpha)\geq0.
 \label{eq:partitionF}
\end{equation}
Validity now follows directly. Bob can gain a negative contribution only from the unique cell containing $\alpha$, and that contribution is $-1$. All other terms are nonnegative. Equality is achieved by setting $b_C=1$ for the containing cell, $b_C=0$ wherever $\ell_C>0$, and either value wherever $\ell_C=0$.

Rewriting $\ind{\alpha_j\neq c_j}=1-\ind{\alpha_j=c_j}$ and averaging over deterministic strategies yields the observable Bell inequality
\begin{equation}
 \ 
 \sum_{C\in\Ccal}\left[
 \sum_{j\in J_C}p(c_j,1|j,C)
 -( |J_C|-1)\pb(1|C)
 \right]\leq1.\ 
 \label{eq:partitionBell}
\end{equation}
Here $C$ is the name of Bob's setting. Every term is an ordinary joint probability or Bob marginal; no simultaneous measurement of Alice's different settings is involved.

\begin{figure}[tbp]
\centering
\begin{tikzpicture}[font=\small]
 \foreach \xx/\head/\col in {0/In the cell/teal,4.7/One mismatch/orange,9.4/Two mismatches/navy}{
   \node[\col,font=\bfseries\small] at (\xx,2) {\head};
 }
 \foreach \xx/\str/\v/\out in {0/{0\,0\,1}/{-1}/{1},4.7/{0\,0\,0}/{0}/{0\text{ or }1},9.4/{0\,1\,0}/{1}/{0}}{
   \node[draw=navy!30,rounded corners,minimum width=3.7cm,minimum height=.7cm] at (\xx,1.15) {$\alpha=(\str)$};
   \node at (\xx,.25) {$\ell_{L_1}(\alpha)=\v$};
   \node[align=center,text width=3.8cm] at (\xx,-.6)
   {Minimizing output:\\[2pt] $b_{L_1}=\out$};
 }
\end{tikzpicture}
\caption{How a cell fixes Bob's minimizing response. For the cell $L_1=\{\alpha:\alpha_2=0,\alpha_3=1\}$ in the binary cube, $\ell_{L_1}=\ind{\alpha_2\neq0}+\ind{\alpha_3\neq1}-1$. A string in the cell gives $-1$, a single mismatch gives a tie, and two mismatches give $+1$. The free coordinate $\alpha_1$ does not enter this coefficient.}
\label{fig:mismatch}
\end{figure}

\subsection{The ties determine each setting's coefficient}
\label{sec:cellties}

We prove property (i) of \cref{prop:facetcriterion} for every cell. Let $h$ be additive and suppose it vanishes on all strings with exactly one mismatch in the fixed coordinates of $C$. These are precisely the strings where $\ell_C=0$.

Write
\begin{equation}
 h(\alpha)=h_0+\sum_j u_j(\alpha_j),
\end{equation}
choosing $u_j(c_j)=0$ on each fixed coordinate and $u_i(0)=0$ on each free coordinate. Such a choice is always possible by moving constants into $h_0$.

First consider a free coordinate $i$. Choose one fixed coordinate to mismatch, keep all other fixed coordinates equal to their prescribed values, and initially set every free coordinate to zero. This is a tie. Varying only $\alpha_i$ leaves it a tie. Since $h$ vanishes before and after the change, subtraction gives $u_i(\alpha_i)-u_i(0)=0$ for every value. Thus the contribution of each free coordinate vanishes identically.

Now fix $j\in J_C$ and choose a tie whose only mismatch is $\alpha_j=a\neq c_j$. The equation $h=0$ gives
\begin{equation}
 h_0+u_j(a)=0.
 \label{eq:celltiecondition}
\end{equation}
This holds for every alternative value $a\neq c_j$, so
\begin{equation}
 u_j(\alpha_j)=-h_0\,\ind{\alpha_j\neq c_j}.
\end{equation}
Consequently,
\begin{equation}
 h(\alpha)=h_0-h_0\sum_{j\in J_C}\ind{\alpha_j\neq c_j}
 =-h_0\ell_C(\alpha).
 \label{eq:celltieunique}
\end{equation}
Every additive function vanishing on the ties is proportional to $\ell_C$, as required. The argument applies equally to a singleton, for which there are no free coordinates.

\subsection{Why the proportionality constants must agree}
\label{sec:cellrelation}

We next prove property (ii). Suppose
\begin{equation}
 \sum_{C\in\Ccal}\lambda_C m_C
 \label{eq:weightedcellminima}
\end{equation}
is additive. By \cref{eq:cellminimum}, its negative is the function
\begin{equation}
 f(\alpha)=\sum_{C\in\Ccal}\lambda_C\ind{\alpha\in C}.
 \label{eq:cellconstant}
\end{equation}
It is constant on each cell, with value $\lambda_C$. Write its additive representation as $f=f_0+\sum_i v_i(\alpha_i)$.

On the line $L_i$, all coordinates except $i$ are fixed, while coordinate $i$ ranges over all $A$ values. Since $f$ is constant on that line, for any $a,a'$ we have
\begin{equation}
 0=f(\alpha_i=a,\text{other coordinates fixed})
   -f(\alpha_i=a',\text{other coordinates fixed})
   =v_i(a)-v_i(a').
 \label{eq:linekills}
\end{equation}
Thus $v_i$ is constant. There is one such line for every $i$, so every single-coordinate contribution is constant. Therefore $f$ is constant on all of $\Omega$. Every cell is nonempty, which forces all $\lambda_C$ to have the same value.

Both conditions of \cref{prop:facetcriterion} are now proved. The inequality in \cref{eq:partitionBell} defines a facet, and every Bob setting is indispensable because $\ell_C=-1$ on its own cell. Since there are $K$ cells, this proves the binary-output lower bound for every $X\geq3$ and $A\geq2$.

This also explains the choice of partition. The partition makes the local bound exact for every Alice string. The coordinate line in each direction ensures that the separate coefficients cannot be rescaled independently while preserving all saturating strategies.

\subsection{A complete five-setting binary example}
\label{sec:fivebinary}

For the cube partition in \cref{eq:cubecells}, write Alice's predetermined bits simply as $\alpha_1,\alpha_2,\alpha_3$. The five functions are
\begin{equation}
 \begin{aligned}
 \ell_1&=\alpha_2-\alpha_3,&
 \ell_2&=\alpha_3-\alpha_1,&
 \ell_3&=\alpha_1-\alpha_2,\\
 \ell_4&=\alpha_1+\alpha_2+\alpha_3-1,&
 \ell_5&=2-\alpha_1-\alpha_2-\alpha_3.
 \end{aligned}
 \label{eq:fiveells}
\end{equation}
The inequality is $1+\sum_{y=1}^5 b_y\ell_y\geq0$. Equivalently,
\begin{equation}
 \sum_{x=1}^3\sum_{y=1}^5M_{xy}p(1,1|x,y)
 +\pb(1|4)-2\pb(1|5)\leq1,
 \qquad
 M=\begin{pmatrix}
 0&1&-1&-1&1\\
 -1&0&1&-1&1\\
 1&-1&0&-1&1
 \end{pmatrix}.
 \label{eq:fiveBell}
\end{equation}
There are no Alice-marginal terms in this representation. Up to exchanging the parties and relabelling settings and outcomes, this facet belongs to the known $I_{5322}$ class~\cite{JesusCruzeiro}, the only class in the complete classification that requires five settings on one side and three on the other.

\begin{table}[htbp]
\centering
\begin{tabular}{c c c c}
\toprule
Alice string $\alpha$ & $(\ell_1,\ell_2,\ell_3,\ell_4,\ell_5)$
 & Saturating $(b_1,b_2,b_3,b_4,b_5)$ & Number\\
\midrule
$000$ & $(0,0,0,-1,2)$ & $(*,*,*,1,0)$ & $8$\\
$001$ & $(-1,1,0,0,1)$ & $(1,0,*,*,0)$ & $4$\\
$010$ & $(1,0,-1,0,1)$ & $(0,*,1,*,0)$ & $4$\\
$011$ & $(0,1,-1,1,0)$ & $(*,0,1,0,*)$ & $4$\\
$100$ & $(0,-1,1,0,1)$ & $(*,1,0,*,0)$ & $4$\\
$101$ & $(-1,0,1,1,0)$ & $(1,*,0,0,*)$ & $4$\\
$110$ & $(1,-1,0,1,0)$ & $(0,1,*,0,*)$ & $4$\\
$111$ & $(0,0,0,2,-1)$ & $(*,*,*,0,1)$ & $8$\\
\bottomrule
\end{tabular}
\caption{Every saturating deterministic strategy of \cref{eq:fiveBell}. An asterisk denotes either binary output, chosen independently of the other asterisks. There are $40$ saturating vertices. The full local polytope has dimension $D=3+5+15=23$, and the proof above establishes that their affine span has dimension $22$. The table makes the output switches used in the tightness proof explicit.}
\label{tab:saturating}
\end{table}

\Cref{tab:saturating} lists all saturating strategies. For example, at $\alpha=001$, settings $3$ and $4$ are tied. Switching either of those outputs while fixing the rest gives two saturating strategies. For any affine Bell expression $F$ of the form \eqref{eq:Fbinary} that vanishes on every deterministic strategy saturating \eqref{eq:fiveBell}, the two switches give $h_3(001)=h_4(001)=0$. Repeating these switches across the table is the concrete version of the argument in \cref{sec:cellties,sec:cellrelation}.

The same prescription gives explicit inequalities in larger scenarios. For five binary Alice settings, the five lines are
\begin{equation}
 \begin{gathered}
 \{00001,10001\},\quad \{10000,11000\},\quad
 \{01000,01100\},\\
 \{00100,00110\},\quad \{00010,00011\}.
 \end{gathered}
\end{equation}
Together with the $22$ uncovered singletons, these give $27$ Bob settings. Equation~\eqref{eq:partitionBell} specifies every coefficient of the resulting tight inequality.

\section{A construction for two Alice settings}
\label{sec:two}

The preceding coordinate-line construction cannot be used at $X=2$. A line varying the first coordinate and a line varying the second intersect in the $A\times A$ grid of response strings. We therefore need a different family.

\subsection{The inequality and its deterministic minimum}

For $\alpha=(\alpha_1,\alpha_2)$, define the indicator functions
\begin{equation}
 r_i=\ind{\alpha_1=i},\qquad c_j=\ind{\alpha_2=j},
 \qquad i,j=1,\ldots,A-1,
\end{equation}
and
\begin{equation}
 r=\sum_{i=1}^{A-1}r_i=\ind{\alpha_1\neq0},
 \qquad
 c=\sum_{j=1}^{A-1}c_j=\ind{\alpha_2\neq0}.
\end{equation}
Bob has a setting $(i,j)$ for each pair of nonzero outcomes and one additional setting, denoted $*$. Thus
\begin{equation}
 Y=(A-1)^2+1=A^2-2(A-1)=K.
\end{equation}
Choose
\begin{equation}
 \ell_{ij}=-r_i+c_j,\qquad \ell_* = r+c-1.
 \label{eq:twoells}
\end{equation}
Each is additive. For $\ell_{ij}$, the value is negative only when $r_i=1$ and $c_j=0$, so
\begin{equation}
 m_{ij}=-r_i(1-c_j)=-r_i+r_ic_j.
 \label{eq:mij}
\end{equation}
For $\ell_*$, the value is negative only when $r=c=0$, giving
\begin{equation}
 m_* = -(1-r)(1-c)=-1+r+c-rc.
 \label{eq:mstar}
\end{equation}
Summing \cref{eq:mij} over $i,j$ gives
\begin{equation}
 \sum_{i,j}m_{ij}=-(A-1)r+rc.
 \label{eq:summedpairminima}
\end{equation}
The terms involving both Alice coordinates cancel when $m_*$ is added:
\begin{equation}
 m_*+\sum_{i,j}m_{ij}=-1-(A-2)r+c.
 \label{eq:twominimasum}
\end{equation}
Thus $g=1+(A-2)r-c$ is additive, and our valid Bell inequality is
\begin{equation}
 F=1+(A-2)r-c
 +\sum_{i,j}b_{ij}(-r_i+c_j)
 +b_*(r+c-1)\geq0.
 \label{eq:twoF}
\end{equation}
As before, for every Alice string, choosing the minimizing Bob responses gives equality.

In observed probabilities this reads
\begin{equation}
\begin{aligned}
 F(p)={}&1+(A-2)\pa(a\neq0|1)-\pa(a\neq0|2)\\
 &+\sum_{i,j=1}^{A-1}
 \bigl[-p(i,1|1,(i,j))+p(j,1|2,(i,j))\bigr]\\
 &+p(a\neq0,1|1,*)+p(a\neq0,1|2,*)-\pb(1|*)\geq0.
\end{aligned}
 \label{eq:twoobserved}
\end{equation}
An event such as $a\neq0$ means the sum over Alice's nonzero outcomes. Equation~\eqref{eq:twoobserved} involves only the two actual Alice settings and is a standard Bell inequality.

\subsection{The ties for the settings \texorpdfstring{$(i,j)$}{(i,j)}}

Every additive function of two Alice answers can be written
\begin{equation}
 h(a_1,a_2)=u(a_1)+v(a_2),
 \label{eq:uv}
\end{equation}
with the constant absorbed into either function. Fix $(i,j)$ and suppose $h$ vanishes wherever $\ell_{ij}=0$. The ties consist of
\begin{equation}
 \{a_1\neq i,\ a_2\neq j\}\ \cup\ \{(i,j)\}.
 \label{eq:ijtie}
\end{equation}
The first set is a nonempty rectangle, even when $A=2$.

Fixing any $a_2\neq j$ and varying $a_1\neq i$ shows that $u$ has a common value $u_0$ outside $i$. Similarly, $v$ has a common value $v_0$ outside $j$. Vanishing on the rectangle implies $u_0+v_0=0$. Let
\begin{equation}
 \delta=u(i)-u_0,\qquad \eta=v(j)-v_0.
\end{equation}
Vanishing at $(i,j)$ then gives $\delta+\eta=0$. Hence, on every string,
\begin{equation}
 h=\delta r_i+\eta c_j
 =\delta(r_i-c_j)=-\delta\ell_{ij}.
 \label{eq:ijtieunique}
\end{equation}
This proves the first tightness condition for all pair settings.

\subsection{The ties for the additional setting}

For $\ell_*=r+c-1$, the ties are exactly the strings
\begin{equation}
 (i,0)\quad\text{and}\quad(0,j),
 \qquad i,j=1,\ldots,A-1.
 \label{eq:starties}
\end{equation}
If \cref{eq:uv} vanishes there, then
\begin{equation}
 u(i)=-v(0)\quad(i\neq0),\qquad
 v(j)=-u(0)\quad(j\neq0).
 \label{eq:startieconstraints}
\end{equation}
Writing $s=u(0)+v(0)$, we find that $h$ equals $s$ at $(0,0)$, zero when exactly one coordinate is nonzero, and $-s$ when both are nonzero. Equivalently,
\begin{equation}
 h=s(1-r-c)=-s\ell_*.
 \label{eq:startieunique}
\end{equation}
The first tightness condition therefore holds at every setting.

\begin{figure}[tbp]
\centering
\begin{subfigure}[t]{.47\textwidth}
\centering
\begin{tikzpicture}[font=\small,scale=.96]
 \foreach \i in {0,1,2}{
 \foreach \j in {0,1,2}{
   \pgfmathtruncatemacro{\lval}{-(\i==1)+(\j==1)}
   \ifnum\lval=0\fill[teal!20] (\j,-\i) rectangle ++(1,-1);\fi
   \node at (\j+.5,-\i-.5) {$\lval$};
 }}
 \draw[step=1,navy!45] (0,0) grid (3,-3);
 \foreach \i in {0,1,2}{
 \node[left] at (0,-\i-.5) {$\i$};
 \node[above] at (\i+.5,0) {$\i$};}
 \node[above=19pt] at (1.5,0) {$\alpha_2$};
 \node[left=19pt] at (0,-1.5) {$\alpha_1$};
 \node[below=8pt] at (1.5,-3) {$\ell_{11}=-r_1+c_1$};
\end{tikzpicture}
\end{subfigure}\hfill
\begin{subfigure}[t]{.47\textwidth}
\centering
\begin{tikzpicture}[font=\small,scale=.96]
 \foreach \i in {0,1,2}{
 \foreach \j in {0,1,2}{
   \pgfmathtruncatemacro{\lval}{(\i>0)+(\j>0)-1}
   \ifnum\lval=0\fill[orange!23] (\j,-\i) rectangle ++(1,-1);\fi
   \node at (\j+.5,-\i-.5) {$\lval$};
 }}
 \draw[step=1,navy!45] (0,0) grid (3,-3);
 \foreach \i in {0,1,2}{
 \node[left] at (0,-\i-.5) {$\i$};
 \node[above] at (\i+.5,0) {$\i$};}
 \node[above=19pt] at (1.5,0) {$\alpha_2$};
 \node[left=19pt] at (0,-1.5) {$\alpha_1$};
 \node[below=8pt] at (1.5,-3) {$\ell_*=r+c-1$};
\end{tikzpicture}
\end{subfigure}
\caption{The two types of tie sets for $X=2$, illustrated with $A=3$. Entries give the value of the indicated coefficient function. Shaded zero entries are the strings where Bob may switch between the two outputs while preserving saturation. Left: the ties form the rectangle excluding row $1$ and column $1$, together with the central point. Right: the ties have exactly one nonzero coordinate. In each panel, an additive function vanishing on the shaded entries is forced to be proportional to the displayed coefficient function.}
\label{fig:twoties}
\end{figure}

\subsection{Four-point differences force all scalars to agree}
\label{sec:mixeddiff}

It remains to prove the second tightness condition. For any function $f$ on the $A\times A$ grid, define
\begin{equation}
 \Delta_{ij}f=f(i,j)-f(i,0)-f(0,j)+f(0,0),
 \qquad i,j=1,\ldots,A-1.
 \label{eq:mixed}
\end{equation}
Every additive function has $\Delta_{ij}f=0$, because its separate row and column contributions cancel. Conversely, if all these differences vanish, then
\begin{equation}
 f(a_1,a_2)=f(a_1,0)+f(0,a_2)-f(0,0)
 \label{eq:mixedconverse}
\end{equation}
for all $a_1,a_2$. This identity is automatic if either coordinate is zero and follows from the vanishing difference otherwise. It explicitly expresses $f$ as an additive function.

For the product indicators,
\begin{equation}
 \Delta_{ij}(r_kc_l)=\delta_{ik}\delta_{jl},
 \label{eq:mixedindicator}
\end{equation}
because they vanish on the reference row and column. Also $\Delta_{ij}(rc)=1$. Applying this to \cref{eq:mij,eq:mstar} gives
\begin{equation}
 \Delta_{ij}m_{kl}=\delta_{ik}\delta_{jl},
 \qquad
 \Delta_{ij}m_*=-1.
 \label{eq:deltaminima}
\end{equation}
Now suppose
\begin{equation}
 f=\sum_{k,l}\lambda_{kl}m_{kl}+\lambda_*m_*
\end{equation}
is additive. Its four-point difference is zero, so \cref{eq:deltaminima} gives
\begin{equation}
 0=\Delta_{ij}f=\lambda_{ij}-\lambda_*
 \quad\text{for every }i,j.
 \label{eq:lambdaequal}
\end{equation}
All the coefficients agree, proving property (ii).

It follows from \cref{prop:facetcriterion} that the inequality in \cref{eq:twoF} defines a facet. Every $\ell_{ij}$ takes the value $-1$ at $(i,0)$, and $\ell_*$ takes the value $-1$ at $(0,0)$. They are all nonzero, so every one of the $(A-1)^2+1=K$ settings is indispensable.

\subsection{Binary and ternary examples}

For $A=2$, there is one pair setting and the additional setting. Writing their outputs $b_1,b_2$, and $r=\alpha_1$, $c=\alpha_2$, gives
\begin{equation}
 F=1-c+b_1(-r+c)+b_2(r+c-1)\geq0.
 \label{eq:CHSHform}
\end{equation}
Introduce signs $A_1=1-2r$, $A_2=1-2c$, and $B_y=1-2b_y$. Direct substitution gives
\begin{equation}
 4F=2-A_1B_1+A_2B_1+A_1B_2+A_2B_2.
\end{equation}
Writing the observed correlators as
\begin{equation}
 E_{xy}=\langle A_x B_y\rangle
       =\sum_{a,b=0}^{1}(-1)^{a+b}p(a,b|x,y),
 \label{eq:observedcorrelators}
\end{equation}
we obtain
\begin{equation}
 E_{11}-E_{21}-E_{12}-E_{22}\leq2,
\end{equation}
which is CHSH up to outcome relabellings.

For $A=3$, Bob has the five settings $11,12,21,22,*$. The deterministic expression is
\begin{equation}
\begin{aligned}
 F={}&1+r-c
 +b_{11}(-r_1+c_1)+b_{12}(-r_1+c_2)\\
 &+b_{21}(-r_2+c_1)+b_{22}(-r_2+c_2)
 +b_*(r+c-1)\geq0.
\end{aligned}
 \label{eq:ternaryfive}
\end{equation}
Here $r=r_1+r_2$ and $c=c_1+c_2$. This inequality defines a facet of the $29$-dimensional local polytope for $(X,Y,A,B)=(2,5,3,2)$, and every Bob setting is indispensable.

\subsection{Complete classification for two ternary Alice settings}
\label{sec:classification}

Consider two three-outcome measurements for Alice and $Y$ binary measurements for Bob. We performed complete facet enumerations with PANDA~\cite{PANDA} for both $Y=5$ and $Y=6$. The input vertices are the deterministic behaviors in \cref{eq:det}: there are $3^2 2^Y$ vertices, namely $288$ and $576$, and the local polytopes have dimensions $29$ and $34$, respectively. Facets are identified up to permutations of each party's settings and independent permutations of the outcomes at each setting. Positivity is included as one class.

The five-setting enumeration gives the ten classes in \cref{tab:ternaryclasses}. To read a row, write $r_i=\ind{\alpha_1=i}$ and $c_j=\ind{\alpha_2=j}$ for $i,j\in\{0,1,2\}$. As above, $r=r_1+r_2$ and $c=c_1+c_2$. The row denotes the inequality
\begin{equation}
  F(\alpha,\bm b)=g(\alpha)+\sum_{y=1}^{m}b_y\ell_y(\alpha)\geq0.
 \label{eq:classrepresentative}
\end{equation}
For example, $b_y r_i$ contributes $p(i,1|1,y)$ and $b_y c_j$ contributes $p(j,1|2,y)$. Single-party terms give the corresponding marginals as in \cref{eq:generalFobserved}. Only the $m$ nonzero setting coefficients are listed. For each of them, switching $b_y$ changes $F$ for some Alice string, so $m$ is the number of indispensable settings.

\begin{center}
\begin{minipage}{\linewidth}
\captionsetup{type=table}
\centering
\small
\setlength{\tabcolsep}{5pt}
\renewcommand{\arraystretch}{1.17}
\begin{tabular}{@{}rrll@{}}
\toprule
Class & $m$ & $g$ & $\ell_1,\ldots,\ell_m$ \\
\midrule
1 & 1 & $0$ & $c_0$ \\
2 & 2 & $c_0$ & $1-r_0-c_0,\; r_0-c_0$ \\
3 & 3 & $2c_0$ & $r_0-c_0,\; r_1-c_0,\; r_2-c_0$ \\
4 & 3 & $2r_1+r_2$ & $1-r_2-c_0,\; r_0-r_1+c_1-c_2,\; r_0-r_1-c_1+c_2$ \\
5 & 3 & $c_0$ & $1-r_0-c_0,\; 1-r_1-c_0,\; 1-r_2-c_0$ \\
6 & 4 & $2c_0+c_1$ & $-r_0+c_2,\; 1-r_1-c_1,\; r_0-r_2-c_0+c_2,\; r_2-c_0$ \\
7 & 4 & $1-r_0$ & $1-r_1-c_0,\; 1-r_2-c_0,\; r_0-c_1,\; r_0-c_2$ \\
8 & 4 & $1$ & $-r_0+c_0,\; -r_1+c_1,\; r_0-c_1,\; r_1-c_0$ \\
9 & 5 & $1-r_0$ & $1-r_1-c_0,\; 1-r_1-c_1,\; 1-r_2-c_0,\; 1-r_2-c_1,\; r_0-c_2$ \\
10 & 5 & $1+r-c$ & $-r_1+c_1,\; -r_1+c_2,\; -r_2+c_1,\; -r_2+c_2,\; r+c-1$ \\
\bottomrule
\end{tabular}
\caption{One representative of each facet class of $(2,5,3,2)$. Each row gives $g+\sum_{y=1}^m b_y\ell_y\geq0$, where $m$ is the number of indispensable Bob settings. Class~1 is positivity; class~2 is a CHSH inequality with Alice's three outcomes grouped into two. Class~10 is the construction in \cref{eq:ternaryfive}.}
\label{tab:ternaryclasses}
\end{minipage}
\end{center}

For $m=1,2,3,4,5$ indispensable Bob settings, the numbers of classes are $1,1,3,3,2$, respectively. The representatives of classes~9 and~10 have $50$ and $64$ saturating deterministic vertices, respectively, so they cannot be related by relabellings. Class~10 is the facet constructed above.

Since $Y_{\sat}(2;3,2)=5$, every facet at six or more settings is obtained by adding unused settings to a representative in \cref{tab:ternaryclasses}. The completed enumeration for $(2,6,3,2)$ also gives ten classes: each is obtained from one of the ten five-setting classes by adding an unused Bob setting. Conversely, every facet for fewer than five settings appears in the five-setting list by input lifting. Restricting a representative to its indispensable settings recovers its smaller-scenario facet, and relabellings preserve this support. Thus the numbers of classes are $1,2,5,8$ for $Y=1,2,3,4$, respectively, and exactly $10$ for every $Y\geq5$. The five classes at $Y=3$ agree with the earlier complete enumeration by Cope and Colbeck~\cite{CopeColbeck}.

\section{Extension to arbitrary Bob outcomes}
\label{sec:Boboutcomes}

Both constructions have now been proved for $B=2$. For each fixed finite $B>2$, replace the binary variable at each setting by
\begin{equation}
 b_y\longmapsto\ind{\beta_y\neq0}.
 \label{eq:outcomelift}
\end{equation}
In the observable inequalities, every event $b=1$ is replaced by $b\neq0$, meaning the sum over outcomes $1,\ldots,B-1$. Validity follows immediately by grouping all nonzero outcomes. Facet preservation under this outcome lifting follows from Pironio's theorem~\cite[Theorem~5]{Pironio}. We give a direct proof for these constructions that also establishes the indispensability of every Bob setting.

On a deterministic behavior specified by $(\alpha,\beta)$, any affine Bell expression $F$ has the form
\begin{equation}
 F(\alpha,\beta)=h(\alpha)+
 \sum_{y=1}^Y\sum_{b=1}^{B-1}\ind{\beta_y=b}\,h_{yb}(\alpha),
 \qquad h,h_{yb}\in\U.
 \label{eq:Fgeneral}
\end{equation}
Uniqueness follows by evaluating first at the all-zero Bob string and then at strings with a single nonzero output.

Let $\widetilde F$ denote the Bell expression obtained by applying the outcome lifting \eqref{eq:outcomelift} to the chosen binary inequality. On a deterministic behavior, its value is
\begin{equation}
 \widetilde F(\alpha,\beta)=g(\alpha)+\sum_{y=1}^Y\ind{\beta_y\neq0}\,\ell_y(\alpha).
 \label{eq:liftedfunctional}
\end{equation}
Let $F$ be any affine Bell expression of the form \eqref{eq:Fgeneral} that vanishes on every deterministic behavior saturating $\widetilde F\geq0$. As in \cref{prop:facetcriterion}, we will show that $F=\lambda\widetilde F$ for some scalar $\lambda$.

At a tie $\ell_y(\alpha)=0$, every outcome at setting $y$ is minimizing. Hold the other settings at minimizing outputs and compare $\beta_y=0$ with $\beta_y=b$. Subtracting the two equations $F=0$ gives $h_{yb}(\alpha)=0$ on the entire tie set. Property (i) of \cref{prop:facetcriterion}, established for both constructions, implies
\begin{equation}
 h_{yb}=\lambda_{yb}\ell_y.
 \label{eq:generalproportional}
\end{equation}
Both constructions satisfy property (ii) of \cref{prop:facetcriterion} and have $Y=K\geq2$. As noted after its proof, for each setting $y$ there is therefore an Alice string $\alpha$ with $\ell_y(\alpha)<0$. At this string, every nonzero Bob outcome gives the same minimum contribution $\ell_y(\alpha)$. Keeping the other settings at minimizing outputs and comparing $\beta_y=b$ with $\beta_y=b'$, where $b,b'\neq0$, yields
\begin{equation}
 0=h_{yb}(\alpha)-h_{yb'}(\alpha)
 =(\lambda_{yb}-\lambda_{yb'})\ell_y(\alpha).
 \label{eq:outcomecomparison}
\end{equation}
The last factor is nonzero, so all $\lambda_{yb}$ at the same setting coincide. Denote the common value by $\lambda_y$.

For each Alice string, choose one nonzero outcome wherever $\ell_y<0$ and choose zero wherever $\ell_y\geq0$. This strategy saturates $\widetilde F\geq0$, so $F(\alpha,\beta)=0$ and hence
\begin{equation}
 h+\sum_y\lambda_y m_y=0.
 \label{eq:liftedvanishing}
\end{equation}
Property (ii), proved separately for the two constructions, forces every $\lambda_y$ to equal a common $\lambda$. Thus $h=\lambda g$ and $h_{yb}=\lambda\ell_y$ for every setting $y$ and every nonzero outcome $b$. Substitution in \eqref{eq:Fgeneral} gives $F=\lambda\widetilde F$ on every deterministic behavior and hence on the normalized nonsignalling affine space. The saturating face again has exactly one independent affine equation, now in the full dimension
\begin{equation}
 D=d+Y(B-1)(d+1).
\end{equation}
It is a facet for arbitrary finite $B\geq2$. Finally, switching a Bob output from zero to a nonzero value at an Alice string with $\ell_y(\alpha)\neq0$ changes $\widetilde F$ by $\ell_y(\alpha)$ while leaving the data at all other settings unchanged. Thus each Bob setting remains indispensable.

We have constructed a facet with $K$ indispensable settings for every $X,A,B\geq2$. Together with the upper bound in \cref{sec:facetupper}, this completes the proof of the setting-saturation identity \cref{eq:main}.

\section{Nonsignalling behaviors that require all \texorpdfstring{$K$}{K} of Bob's settings to reveal nonlocality}
\label{sec:sharp}

We now show that a facet with $K$ indispensable Bob settings yields a nonlocal nonsignalling behavior whose restriction to every nonempty proper subset of Bob settings is local.

Work in the $(X,K,A,B)$ scenario and let $F(p)\geq0$ be one of the constructed tight Bell inequalities. Denote its facet by $\mathcal T=\{p\in\Lloc^{A,B}_{X,K}:F(p)=0\}$, and choose $p^\circ$ in the relative interior of $\mathcal T$. Such a point exists; for example, the average of all the vertices of $\mathcal T$ with strictly positive weights lies in its relative interior. If a proper subface contained this average, a supporting inequality for that subface would vanish on every vertex of $\mathcal T$, and hence on the whole facet, a contradiction.

For every nonempty proper subset $S$ of Bob settings, take a finite facet description of $\Lloc^{A,B}_{X,|S|}$ and lift it to the full scenario. There are finitely many such inequalities. Each of them is \emph{strictly} satisfied by $p^\circ$. To see this, suppose a lifted inequality were saturated. The relative-interior argument in \cref{sec:facetupper} would force its supporting face to contain $\mathcal T$. This face is proper: a restricted local model that satisfies the inequality strictly extends to a full local model by fixing the omitted Bob outputs. It would therefore equal $\mathcal T$, expressing $F$ as an inequality that omits at least one setting. This contradicts indispensability.

The point $p^\circ$ also has strictly positive probability entries. Indeed, each positivity inequality is valid and proper. If an entry vanished at $p^\circ$, the same relative-interior argument would place the whole facet in that positivity hyperplane. Their affine equations would then be proportional. But a positivity inequality uses only one Bob setting, while $F$ has $K\geq2$ indispensable settings. This is impossible.

Choose a local point $p^1$ with $F(p^1)>0$, which exists because $F$ defines a proper facet. Move a small distance outwards through $p^\circ$:
\begin{equation}
 p^\varepsilon=p^\circ+\varepsilon(p^\circ-p^1),
 \qquad \varepsilon>0.
 \label{eq:outward}
\end{equation}
This affine combination preserves normalization and nonsignalling, and
\begin{equation}
 F(p^\varepsilon)=-\varepsilon F(p^1)<0.
 \label{eq:violateoutward}
\end{equation}
Thus $p^\varepsilon$ lies outside the local polytope. It remains to choose $\varepsilon$ so that $p^\varepsilon$ is a nonsignalling behavior with local restrictions.

For sufficiently small $\varepsilon$, every entry remains positive, and every lifted inequality from a proper subset remains satisfied. Since there are finitely many lifted inequalities and each is strictly satisfied at $p^\circ$, continuity ensures that they remain satisfied for sufficiently small $\varepsilon$. More explicitly, for any one of these affine quantities $h$ with $h(p^\circ)>0$, the only restriction arises if $h(p^1)>h(p^\circ)$; then it is enough to choose
\begin{equation}
 0<\varepsilon<
 \frac{h(p^\circ)}{h(p^1)-h(p^\circ)}.
 \label{eq:epsilon}
\end{equation}
Choose $\varepsilon>0$ smaller than all applicable bounds, including those for the probability entries. Then $p^\varepsilon$ is a nonlocal nonsignalling behavior and every proper Bob-setting restriction is local.

This proves the optimality assertion in the third part of \cref{thm:main}. The construction establishes sharpness in the nonsignalling set.

\section{Consequences and discussion}
\label{sec:discussion}

The exact cutoff gives a finite endpoint for the classification of tight Bell inequalities when Alice's scenario and Bob's outcome alphabet are fixed. Beyond $K$, every facet is obtained by ignoring some Bob settings. The cutoff may still be large: it grows exponentially with $X$.

\begin{table}[tbp]
\centering
\begin{tabular}{c rr}
\toprule
Alice settings $X$ & Binary outcomes $A=2$ & Ternary outcomes $A=3$\\
\midrule
2&2&5\\
3&5&21\\
4&12&73\\
5&27&233\\
6&58&717\\
\bottomrule
\end{tabular}
\caption{Exact values of $Y_{\sat}(X;A,B)=A^X-X(A-1)$. Every entry holds for all finite $B\geq2$. The lower-bound constructions give a tight inequality with exactly the displayed number of indispensable Bob settings.}
\label{tab:cutoffs}
\end{table}

For binary Alice outcomes, the first values are $2,5,12,27,58$. The two-setting case recovers CHSH. With three binary Alice settings, no facet can require more than five Bob settings. For $B=2$, the complete classification of $(5,3,2,2)$ therefore extends to every $(N,3,2,2)$ with $N\geq5$, after exchanging the parties. This proves the conjecture of Jesus and Cruzeiro that the six nontrivial facet classes suffice for all such $N$~\cite{JesusCruzeiro}. The cube construction gives an explicit facet requiring all five Bob settings. For ternary Alice outcomes, the first cutoffs are $5,21,73,233,717$.

There is also a direct use for locality testing. A nonlocal nonsignalling behavior always has a nonlocal restriction to at most $K$ Bob settings. For a particular behavior, the proof gives the possibly smaller bound $\dim Q(\pa)+1$. Zero Alice marginals may reduce this dimension. This bounds the size of the subsets that must be checked, although the number of such subsets may still be large.

The two parts of the proof use the same response-string geometry in different ways. In the upper bound, fixing Alice's one-setting marginals removes the constant and single-setting degrees of freedom from a distribution over strings. In the lower bound, those same additive functions are exactly the functions allowed in a standard Bell functional at one Bob setting. The construction arranges the saturating strategies so that these functions leave only one independent affine equation.

The result concerns the full local polytope, including marginal probabilities. A projection retaining only correlators is a different polytope and has its own facet structure. Likewise, indispensability for a facet is distinct from indispensability for the violation by a specified quantum state. The upper bound $K$ applies to quantum correlations as well. For each fixed $X,A,B\geq2$, does the $(X,K,A,B)$ scenario admit a nonlocal quantum behavior that becomes local whenever any one of Bob's settings is removed? Our nonsignalling construction does not settle this question in general.

\section*{Acknowledgements}

This work was funded by national funds through FCT---Fundação para a
Ciência e a Tecnologia, I.P., and, when eligible, co-funded by European
Union funds under project UID/50008/2025---Instituto de Telecomunicações,
DOI: \href{https://doi.org/10.54499/UID/50008/2025}{10.54499/UID/50008/2025}.

OpenAI ChatGPT (GPT-6 Astra) was used to assist with exact algebraic and computational cross-checks, the exploration and refinement of candidate
mathematical arguments, and manuscript organization and editing. The authors revised all AI-assisted material and take full responsibility for the content of the manuscript.

\end{document}